\documentclass[12pt,onecolumn]{revtex4-2}
\usepackage[T1]{fontenc}
\usepackage{amsmath}
\usepackage{textcomp}
\usepackage[latin1]{inputenc}
\usepackage{graphicx,bm}
\usepackage{blindtext, rotating}
\usepackage{mathtools}
\usepackage{gensymb}
\usepackage{enumitem}
\usepackage{xcolor}
\usepackage{babel}
\usepackage{hyperref}
\usepackage{braket}
\usepackage{natbib}
\usepackage{tabularx}
\usepackage{amssymb}
\usepackage{verbatim}

\makeatletter
\renewcommand{\p@subsection}{}
\makeatother

\begin{document}
	
	\title{Evidence of linear and cubic Rashba effect in non-magnetic heterostructure}
	%
	\author{Sanchari Bhattacharya}
	\email{bh.sanchari@gmail.com}
	\affiliation{Department of Physics and Astronomy, National Institute of Technology, Rourkela, Odisha, India, 769008}
	\author{Sanjoy Datta}
	\email{dattas@nitrkl.ac.in}
	\affiliation{Department of Physics and Astronomy, National Institute of Technology, Rourkela, Odisha, India, 769008}
	
	\begin{abstract} 
		The $\text{LaAlO}_\text{3}$/$\text{KTaO}_\text{3}$ system serves as a prototype to study the electronic properties that 
		emerge as a result of spin-orbit coupling. In this article, we have used first-principles calculations to systematically study two types of 
		defect-free (0 0 1) interfaces, which are termed as Type-I and Type-II. 
		While the Type-I heterostructure produces a two dimensional electron gas, the Type-II heterostructure hosts an oxygen-rich two dimensional hole gas
		at the interface. Furthermore, in the presence of intrinsic spin-orbit coupling, we have found evidence of both cubic and linear Rashba interactions
		in the conduction bands of the Type-I heterostructure. On the contrary, there is spin-splitting of both the valence and the conduction bands in
		the Type-II interface, which are found to be only linear Rashba type. 
		Interestingly, the Type-II interface also harbours a potential photocurrent transition path, making it an excellent platform to study 
		the circularly polarized photogalvanic effect.
	\end{abstract}
	\maketitle
	\section{Introduction}
	Spin based electronics has taken the lead in the fast expanding field of quantum technologies for carrying information by spin instead of charge~\cite{hirohata2020review}. These energy-efficient devices necessarily require manipulation of local spins for efficient processing of 
	quantum information. In recent years, manipulation of the spin degree of freedom in solid-state materials via spin-orbit coupling, especially 
	the Rashba spin-orbit coupling, has emerged as one of the most promising approach for developing such next generation energy-efficient 
	electronic devices~\cite{bercioux2015quantum,manipatruni2019scalable}. Dresselhaus and Rashba were the first to observe that the combined 
	effect of the intrinsic spin-orbit coupling (SOC) and the bulk inversion asymmetry could lead to spin-splitted energy bands in non-centrosymmetric 
	zinc-blende or wurtzite semiconductors~\cite{dresselhaus1955spin,rashba1960properties}. Later, Bychkov and Rashba ~\cite{bychkov1984properties} 
	discovered that spin-splitted energy bands can also appear in  two-dimensional materials due to structural 
	inversion asymmetry of the confining potential, which is now commonly referred as Rashba spin-orbit (RSO) interaction. The advantage of controlling 
	the RSO interaction in a material by an external electric field holds the promise to resolve the designing issues of the spintronic devices which 
	typically arises due to the inclusion of local magnetic field~\cite{awschalom2009spintronics, Dil_2009}. The motivation of designing spin-based 
	electronic devices without the ferromagnetic elements has given rise to the rapidly emerging field of spin-orbitronics~\cite{soumyanarayanan2016emergent,trier2022oxide}. 
	The most effective method of creating spin current inside the non-magnetic materials is by spin-charge interconversion. 
	The electrically induced regulation of the spin dynamics through spin-orbit coupling is the most practical and desirable 
	method of achieving this interconversion~\cite{vaz2019mapping, manchon2015new}. 
	The principle of spin-orbitronics was first explored in semiconductors and metals, though within some years oxide-heterostructures 
	have emerged to be the most promising platform with a plethora of unique characteristics~\cite{bibes2011ultrathin}.
	Complex oxide heterostructures are widely studied for their numerous fascinating properties, such as 2D superconductivity, 
	coexistence of ferromagnetism \& superconductivity, colossal magnetoresistance etc.~\cite{brinkman2007magnetic,Pentcheva_2010,hwang2012emergent,spaldin-dft-interface-11}, and their promising applications in the manufacturing of next generation all-oxide solid state devices~\cite{ohno1999electrical,he2021synthesis}. The interface of a heterostructure made of complex oxide perovskites breaks the structural 
	inversion symmetry and therefore the electron/hole gas confined at the interface experiences a potential gradient perpendicular to the 
	conduction plane~\cite{winkler-2000}. Oxide interfaces offer a flexible platform for creating, controlling, and detecting spin currents or spin 
	textures. 
	$\text{SrTiO}_\text{3}$ based surfaces and interfaces are first observed as the promising candidate for creating 
	two dimensional electron gas (2DEG) with large trasnsport properties and highly confined quantum well~\cite{ohtomo2004high,popovic-prl-08,santander2011-SrTiO3}. As $\text{SrTiO}_\text{3}$ has strongly correlated 3d-orbital, the effect of intrinsic SOC is expected to be strong in those surfaces and interfaces making the system a suitable platform for generating Rashba spin splitting~\cite{soc-2013-sto}. The first evidence for RSO interaction in oxide surfaces 
	indeed emerged in the 2DEG generated at the $\text{SrTiO}_\text{3}$ surface along the (0 0 1) direction~\cite{nakamura2012experimental}.
	Irrespective of various interesting SO-based properties in $\text{LaAlO}_\text{3}$/$\text{SrTiO}_\text{3}$ heterostructure,
	one of the main difficulty with this system is the presence of ferromagnetism at the interface, and the RSO interaction is dependent on 
	the magneto-resistance tuning~\cite{kong2021tunable}.
	
	To overcome this situation, there is an ongoing effort to create oxide heterostructure with non-magnetic interface and stronger SOC, and
	$\text{KTaO}_\text{3}$ has emerged as an exciting alternative for its novel electronic as well as spintronic properties due to the presence of 
	the 5d-orbital and high Z-value.~\cite{gabay-2010,hwang2012emergent,gupta2022ktao3}. The carrier dependence of spin precision length is shorter in 
	$\text{KTaO}_\text{3}$ than that in $\text{SrTiO}_\text{3}$ ~\cite{nakamura2012experimental}. 
	Due to the polar nature of $\text{KTaO}_\text{3}$, the surface with $\text{TaO}_\text{2}^{+}$ or $\text{KO}^{-}$ 
	termination gives rise to naturally induced 2DEG and its counterpart two-dimensional hole gas (2DHG) 
	at the interfaces~\cite{wang2016creating,2DHG-JMCC-20}. $\text{KTaO}_\text{3}$, being a polar material, has wide range of properties to be used as 
	a substrate material for its novel surface properties~\cite{li2003surface,wang2022surface}. The in-built electric field at the surface of $\text{KTaO}_\text{3}$, provides an ideal system for studying RSO~\cite{kto-surface-rso} and recently experimental evidences of spin splitting have been found in the surface bands~\cite{2012-kto-soc,2012-orbital-kto}. In comparison to $\text{LaAlO}_\text{3}$/$\text{SrTiO}_\text{3}$ heterostructure, $\text{LaAlO}_\text{3}$/$\text{KTaO}_\text{3}$
	has larger charge carrier density at interface with high electron mobility as $\text{LaAlO}_\text{3}$ is also a polar material~\cite{zhang2017highly}.
	Besides, in the bulk state, $\text{KTaO}_\text{3}$ and $\text{LaAlO}_\text{3}$ are cubic with space group Pm$\bar{3}$m (No.221). 
	Due to the comparable band gaps and lattice parameters, $\text{LaAlO}_\text{3}$ and $\text{KTaO}_\text{3}$ are considered
	as very good candidates for creating heterostructure with an effective charge accumulation at the interfaces. Recently, there have been 
	theoretical and experimental studies of $\text{LaAlO}_\text{3}$/$\text{KTaO}_\text{3}$ heterostructures~\cite{high-mobility-kto,varotto2022direct}. 
	However, detailed first-principle study on the effect of intrinsic SOC and the possibility of RSO interaction has remained unexplored.
	
	Motivated by this, in this article, we have investigated the electronic and spintronics properties of the $\text{LaAlO}_\text{3}$/$\text{KTaO}_\text{3}$ heterostructure for two possible interfaces, which are $\text{TaO}_\text{2}^{+}$/$\text{LaO}^{+}$ (Type-I) and 
	$\text{KO}^{-}$/$\text{AlO}_\text{2}^{-}$ (Type-II) using density functional theory (DFT) based first-principle calculations.
	We have found 2DEG in Type-I heterostructure whereas Type-II system exhibit 2DHG in its interface. In the presence of SOC, Type-I system produce both the cubic and linear RSO splitting which has been originated from conduction bands. On the contrary, Type-II heterostructure predominantly produce large k-linear RSO coupling strength which has been produced from both the conduction bands and valence bands. As earlier reported in Ref.~\cite{kto-cpge-rso}, the production of circularly polarized photocurrent is possible in $\text{KTaO}_\text{3}$ slab system where giant RSO interaction is present. Following this, we have found a possible photocurrent transition route for the Type-II system.
	
	In the present report, first we have discussed  structural and computational details in section~\ref{Computational and Structural Details}. 
	In Section~\ref{effective hamiltonian} we have discussed on the effective Hamiltonian for the $\rm{C_{4v}}$ little point group.
	Following this we have systematically reported the Type-I and Type-II system results in Section~\ref{Results}. The orbital contribution in the absence of SOC has been discussed
	in subsections ~\ref{t1-nso} and~\ref{t2-nso}. The explanation of RSO and spin texture for both type systems is covered in sections.~\ref{t1-rso} and~\ref{t2-rso}.
	In section ~\ref{Conlusion}, we finally come to a conclusion.
	\section{Structural and Computational Details}\label{Computational and Structural Details}
	\begin{figure*}[htbp!]
		\centering
		\includegraphics[width=\textwidth,trim={0cm 13cm 0cm 7cm},clip=true]{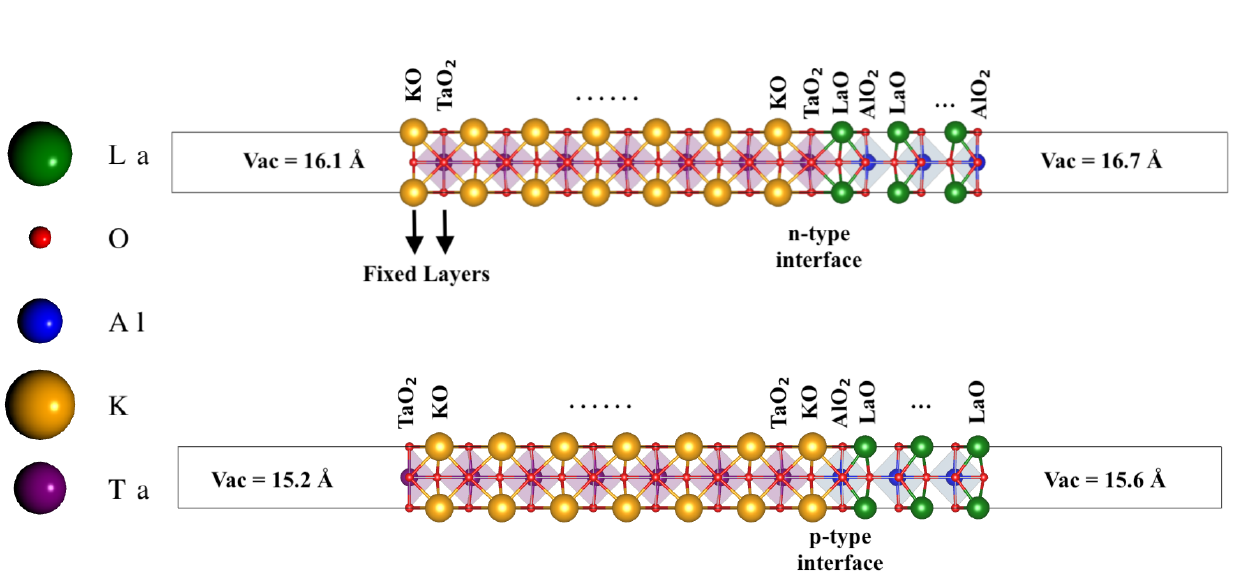}\vspace{-5.5mm}
		\caption{The slab model for the $\text{(KTO)}_\text{6.5}$/$\text{(LAO)}_\text{2.5}$ heterostructure with 
			($\text{TaO}_\text{2}^{+}$/$\text{LaO}^{+}$) and ($\text{KO}^{-}$/$\text{AlO}_\text{2}^{-}$) interface.} 
		\label{Fig:structure}
	\end{figure*}
	In Fig.~\ref{Fig:structure}, we present the asymmetric slab model along (0 0 1) made of $\text{(KTO)}_\text{6.5}$ and $\text{(LAO)}_\text{2.5}$
	with two types of defect-less interfaces. No strain has been applied in the system.
	The perovskite oxide material with parent formula $\rm{ABO_3}$ can be stacked as $\rm{BO_2}$ and $\rm{AO}$. 
	$\text{KTaO}_\text{3}$ substrate comprises of alternative $\text{KO}^{-}$ and ${\text{TaO}_\text{2}}^{+}$ layer,
	whereas $\text{LaAlO}_\text{3}$ epitaxial layers are made of $\text{LAO}^{+}$ and ${\text{AlO}_\text{2}}^{-}$ sub-layers.
	To maintain the interface stoichiometry there are two possible kind of interfaces in the heterostructure created by $\text{LaAlO}_\text{3}$ and $\text{KTaO}_\text{3}$ along (0 0 1) direction, i.e., $\text{TaO}_\text{2}^{+}$/$\text{LaO}^{+}$ and $\text{KO}^{-}$/$\text{AlO}_\text{2}^{-}$. 
	Due to the asymmetry present in the heterostructures, both the systems correspond to P4mm (No. 99) space group,
	which does not have spatial inversion symmetry or the mirror symmetry.
	
	The first-principle density functional calculations are performed by Quantum $\it{Espresso}$ (QE) package~\cite{qe}. 
	The exchange-correlation of the electron interactions are taken into account by Perdew-Burke-Ernzerhof (PBE) functional~\cite{pbe}. 
	The projector augmented wave (PAW) basis set has been used to include the interaction between valence electrons and core ions~\cite{paw}.
	The on site Coulomb interactions of Ta-5d orbitals are considered by using the standard PBE+U method~\cite{ldau}.
	The effective value $\text{U}_\text{eff}=3$ eV is employed for Ta-5d states in this work, as it is well established that such a value is appropriate to 
	describe the strongly-correlated states of KTO~\cite{xu-pccp-21}. Using the linear response theory implemented in the QE package 
	we have independently verified the $U$ value of the Ta-5d states for our heterostructures. 
	The plane-wave basis with a cut-off energy of 75 Ry is used to expand the electronic wavefunctions. 
	Full geometry optimizations are performed using the quick-min Verlet damped dynamics algorithm~\cite{damp-optimization}.
	Here, $\Gamma$-centered k-point grids for sampling~\cite{monkhorst1976special} the first Brillouin zone are set to 
	$8 \times 8 \times 1$ for $1 \times 1 \times (m+n)$ slab models of LAO/KTO heterostructures where m and n are the number of sublayers
	of $\text{KTaO}_\text{3}$ and $\text{LaAlO}_\text{3}$.
	Scalar relativistic effect has been taken for optimization of the structure and later on fully relativistic effect
	has been included in selected pseudopotentials to study the SOC effect. 
	It has been tested that there is no significant effect of SOC in the structure optimization.
	As shown in Fig.~\ref{Fig:structure}, in the Type-I heterostructure the bottom two atomic sublayers $\text{KO}$ and 
	$\text{TaO}_\text{2}$ are kept fixed during structural optimization, while in the case of 
	the Type-II heterostructure no layer has been kept fixed.
	To minimize the interaction between neighboring slabs, a vacuum layer with a thickness of $\geq$ 30~\AA~is 
	applied along the z direction (out of plane).
	All atoms except those in the fixed sublayer/s are fully relaxed until the force 
	acting on each atom is < $10^{-3}$ Ry/Bohr. 
	The convergence criteria of the total energy is set to be < $10^{-4}$ Ry/Bohr.
	Spin-polarized calculations has been done on the slab system to check the collinear
	magnetism, which has been found to be zero, and hence all the SOC calculations have been conducted
	in the slab system by setting the initial magnetism on Ta as 0 $\mu_B$.
	The experimental lattice parameters of bulk $\text{LaAlO}_\text{3}$ and $\text{KTaO}_\text{3}$ are 3.790~\AA~and 
	3.989~\AA~\cite{LaAlO3-lattice}, whereas by using PBE type pseudopotential we have found the theoretical lattice parameters of 
	KTO and LAO to be 4.02~\AA~and 3.81~\AA, which has also been reported earlier~\cite{2DHG-JMCC-20}. 
	For the heterostructures, we use an average lattice constant of 4.02~\AA~for both the KTO and LAO regions.
	The theoretical band gaps of LAO and KTO with PBE pseudopotential are 3.61 eV and 2.84 eV which are comparable with the experimental 
	band gaps 5.6 eV and 3.5 eV respectively~\cite{kto-band-gap}.
	Unlike $\text{LaAlO}_\text{3}$/$\text{SrTiO}_\text{3}$ heterostructure, there is no critical thickness of epitaxial layers for insulator-to-metal transition in $\text{LaAlO}_\text{3}$/$\text{KTaO}_\text{3}$ heterostructure~\cite{2DHG-JMCC-20}.
	
	\section{Effective $\bold{k.p}$ Hamiltonian}\label{effective hamiltonian}
	The Rashba effect is a momentum-dependent spin-splitting of an energy band resulting from the combined effect of intrinsic spin-orbit interaction and broken inversion symmetry. Both the heterostructures considered in this work have $c_{4v}$ point group symmetry.  
	The little group has three high symmetry points when it is considered for 2D materials, which are $\Gamma$ (0,0,0), X (0.5,0,0), M(0.5,0.5,0).
	In these systems, the splitted bands can be described by an effective two band Hamiltonian~\cite{hamiltonian-c4v-vajna} at $\Gamma$ point given by,
	\begin{equation}
		\begin{split}
			H_{c_{4v}} = \alpha ({k_x}^{2}+{k_y}^{2}) + \beta {k_z}^{2} +\alpha_{R1} ({k_x}{\sigma_{y}}-{k_y}{\sigma_{x}}) \\
			+ \alpha_{R2}{k_x}{k_y}({k_y}{\sigma_{y}}-{k_x}{\sigma_{x}})+ \alpha_{R3}({{k_x}^{3}}{\sigma_{y}}-{{k_y}^{3}}{\sigma_{x}}),
		\end{split}
		\label{hamiltonian}
	\end{equation}
	where $\alpha_{R1} ({k_x}{\sigma_{y}}-{k_y}{\sigma_{x}})$ is the linear RSO interaction term, whereas $\alpha_{R2}$ and $\alpha_{R3}$
	are the coefficient of cubic RSO interaction. To estimate the strengths of the linear and cubic RSO interactions in our systems, 
	we consider the $\Gamma-\rm{X}$ high symmetry path. Imposing this condition we solve Eq.~\ref{hamiltonian} to obtain the following two 
	eigenvalue equations given by,
	\begin{equation*}
		\epsilon_{1} = \alpha k_{x}^{2} + \alpha_{R1} k_{x} + \alpha_{R3} k_{x}^{3}
	\end{equation*}
	\begin{equation*}
		\epsilon_{2} = \alpha k_{x}^{2} - \alpha_{R1} k_{x} - \alpha_{R3} k_{x}^{3}.
	\end{equation*}
	The amount of momentum dependent spin-splitting $\Delta_{R}$ is given by,
	\begin{equation}
		\Delta_{R} = \epsilon_{1} - \epsilon_{2} = 2 \alpha_{R1} k_{x} + 2 \alpha_{R3} k_{x}^{3}
		\label{fitting}
	\end{equation}
	The linear and cubic coupling strengths are obtained by fitting Eq.~\ref{fitting} to the DFT data.  
	\section{Results}\label{Results}
	In this section we discuss the aforementioned two types of interface made of  $\text{LaAlO}_\text{3}$ and $\text{KTaO}_\text{3}$. First we present the results of $\text{TaO}_\text{2}^{+}$/$\text{LaO}^{+}$ (type-I) interface which will be followed by $\text{KO}^{-}$/$\text{AlO}_\text{2}^{-}$ (type-II). The role of SOC and the origin of Rashba interactions are presented along with each of the interface.
	\subsection{Type-I heterostructure without SOC}\label{t1-nso}
	\begin{figure}[htbp!]
		\includegraphics[height=0.3\textheight,width=0.5\textwidth,trim={0.45cm 1cm 0.3cm 0.55cm},clip=true]{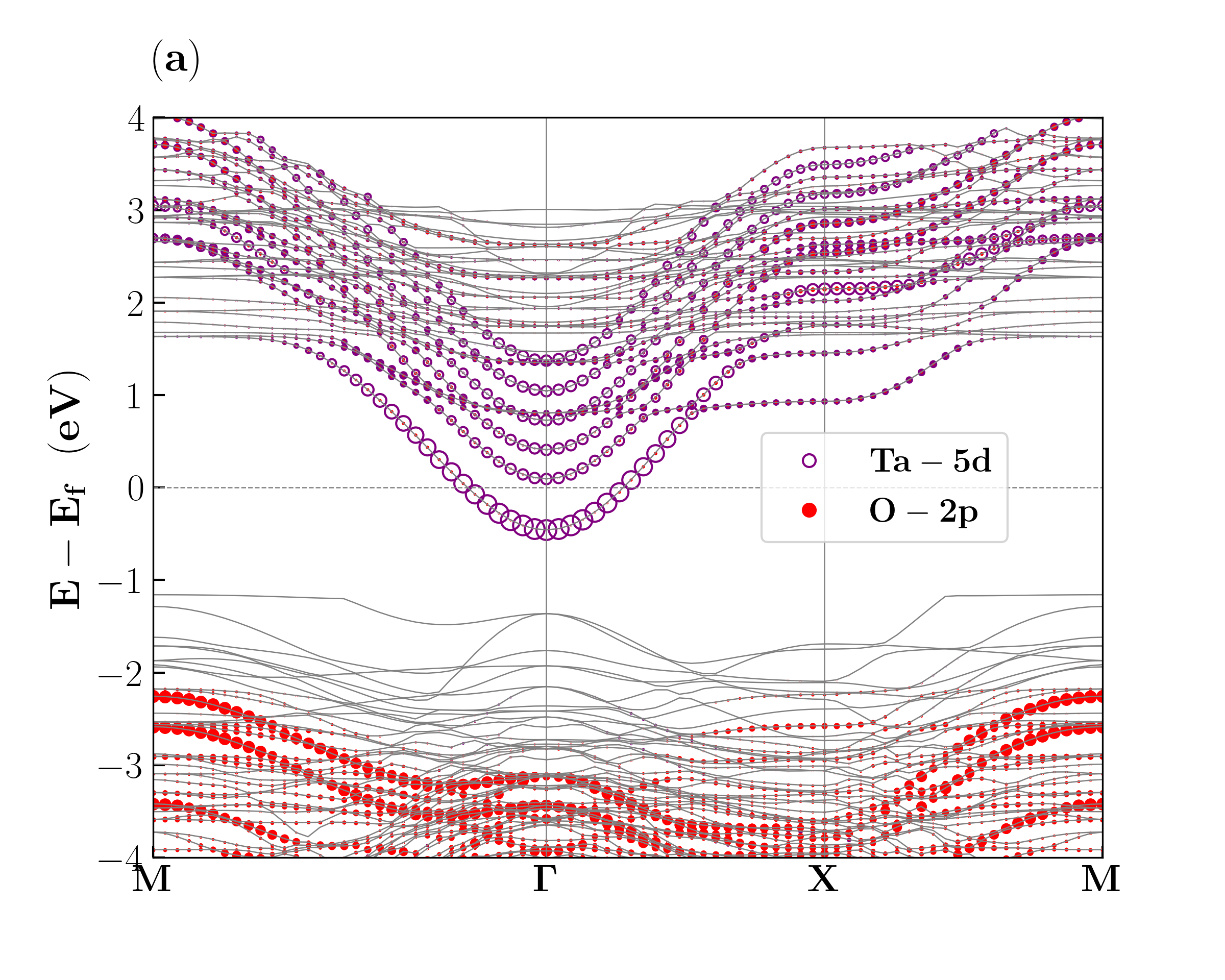}\hspace{-0.6cm}
		\includegraphics[height=0.3\textheight,width=0.5\textwidth,trim={1cm 1cm 1cm 0.55cm},clip=true]{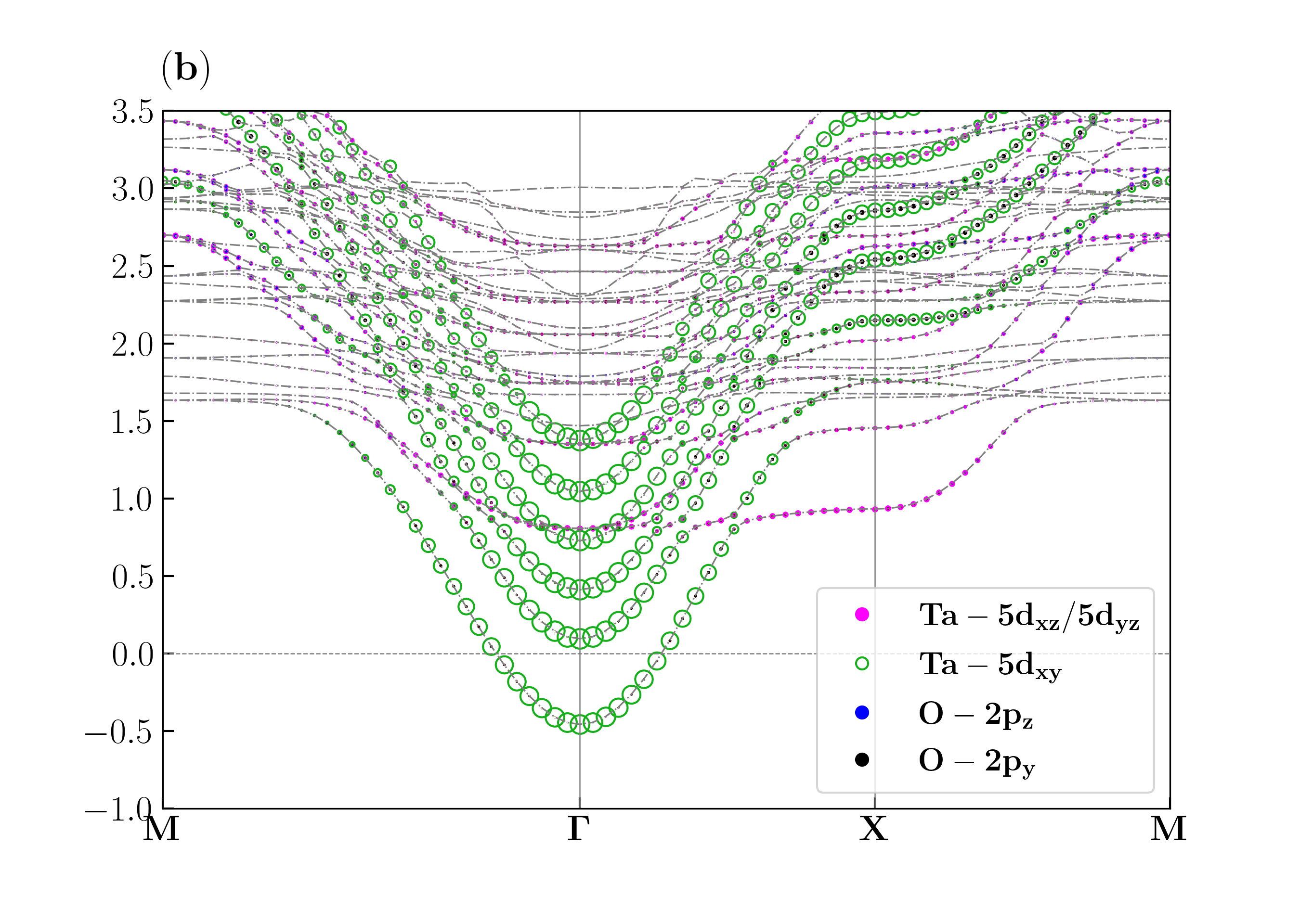} 
		\includegraphics[height=0.67\textheight,width=0.55\textwidth,trim={0.3cm 1.4cm 1cm 3cm},clip=true]{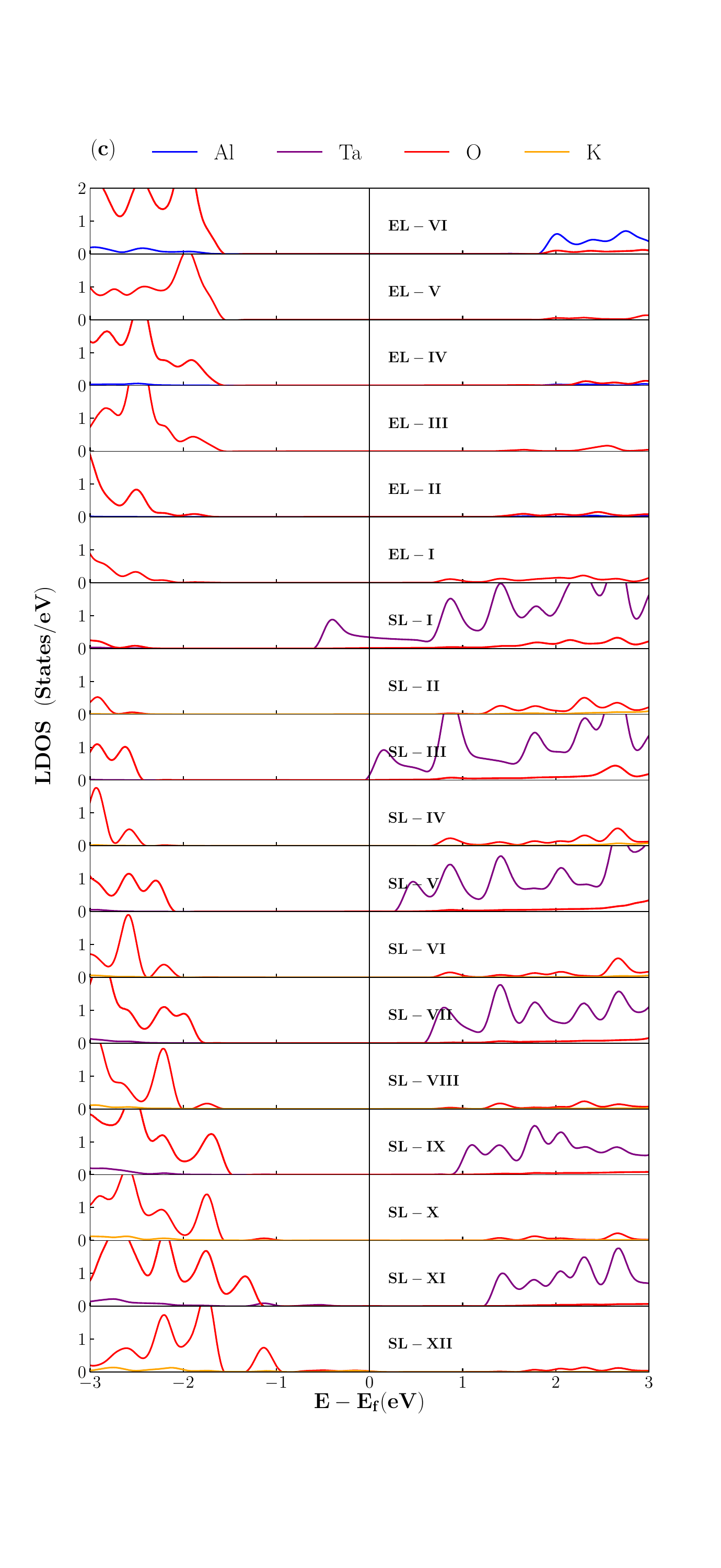}\vspace{-1cm}
		\caption{(a) Band structure of the Type-I heterostructure. Grey bands represent the states originating from fixed substrate layers,
			(b) Zoomed-in display of the subbands coming from Ta-5d orbitals, and (c) Layer-resolved density of states for the slab system. Fixed
			substrate layers are not shown in the picture.}
		\label{Fig:ntype-band}
	\end{figure} 
	The $\text{TaO}_\text{2}^{+}$/$\text{LaO}^{+}$ interface is made by two polar end of parent materials
	which helps the accumulation of electron-like carriers at the interface, thereby giving rise to a 2DEG. 
	Fig.~\ref{Fig:ntype-band}(a) presents the band structure in the absence of SOC along $\rm{M}-\Gamma-\rm{X}-\rm{M}$ high symmetry path. 
	The band structure clearly shows that close to the Fermi energy the atomic contributions are mainly due to the Ta-5d orbitals. However, above the 
	Fermi energy at about $1~\rm{eV}$ or more, O-2p orbitals also contribute, albeit tiny, towards the conduction bands. 
	However O-2p orbitals contribute significantly to form the valence bands. The zoomed-in view of the orbital contributions has been presented in~Fig.~\ref{Fig:ntype-band}(b).
	
	Due to the crystal field splitting (CFS) the degeneracy of the $\rm{t_{2g}}$ orbitals of Ta-5d are lifted 
	into a $\rm{d_{xy}}$ singlet and a $\rm{d_{yz}}$/$\rm{d_{xz}}$ doublet. At the interface the $\rm{d_{xy}}$ band is lower in energy than 
	the $\rm{d_{yz}}$ or $\rm{d_{xz}}$ band with a parabolic structure centered around the $\Gamma$ point.
	Each $\rm{d_{xy}}$ band originates from individual sublayer (marked by open green circle in Fig.~\ref{Fig:ntype-band}(b)),
	whereas all the sublayers contribute to form each of the $\rm{d_{yz}}$/$\rm{d_{xz}}$ bands (marked by solid magenta circle in 
	Fig.~\ref{Fig:ntype-band}(b)). $\rm{O-2p_y}$ orbital hybridizes with the $\rm{d_{yz}}$/$\rm{d_{xz}}$ orbitals.  
	The angular momentum of $\rm{d_{yz}}$ and $\rm{d_{xz}}$ are same with opposite orientation hence the occupation of 
	$\rm{d_{yz}}$ and $\rm{d_{xz}}$ orbitals are equal in weightage. The layer-resolved density of states (LRDOS) without spin-orbit interaction 
	are presented in Fig.~\ref{Fig:ntype-band}(c), which reconfirms that the Ta-5d orbitals are predominantly present at the interface that 
	gives rise to the 2DEG. 
	
	A visible inter orbital crossing takes place between $\rm{d_{xy}}$ and $\rm{d_{yz}}$/$\rm{d_{xz}}$ in the $\Gamma-\rm{X}$ path.
	The inter-orbital crossing between $\rm{d_{xy}}$ and $\rm{d_{yz}}$/$\rm{d_{xz}}$ bands gives the usual multiorbital (M-O) effect, which has been 
	reported earlier at the interface of $\text{SrTiO}_\text{3}$/$\text{LaAlO}_\text{3}$ heterostructure~\cite{soc-2013-sto,13-khalsa-MO,2014quasiparticle-MO,kong2021tunable}.
	However, in contrast to this earlier reported results, in the case of $\text{KTaO}_\text{3}$/$\text{LaAlO}_\text{3}$ heterostructure 
	the M-O effect is not confined  within the d-orbitals. Due to the CFS, $\rm{O-2p_y}$ orbital also takes part along with the d 
	orbitals in this kind of M-O effect. We have found that the $\rm{p_y}$ orbital hybridizes with the degenerate $\rm{d_{yz}}$/$\rm{d_{xz}}$ 
	orbitals, which has not been reported earlier for $\text{KTaO}_\text{3}$-based 2DEG. 
	\subsection{Type-I heterostructure with SOC}\label{t1-rso}
	\begin{figure}[htbp!]
		\includegraphics[width=0.5\textwidth,height=0.3\textheight,trim={0.3cm 1cm 1cm 0.65cm},clip=true]{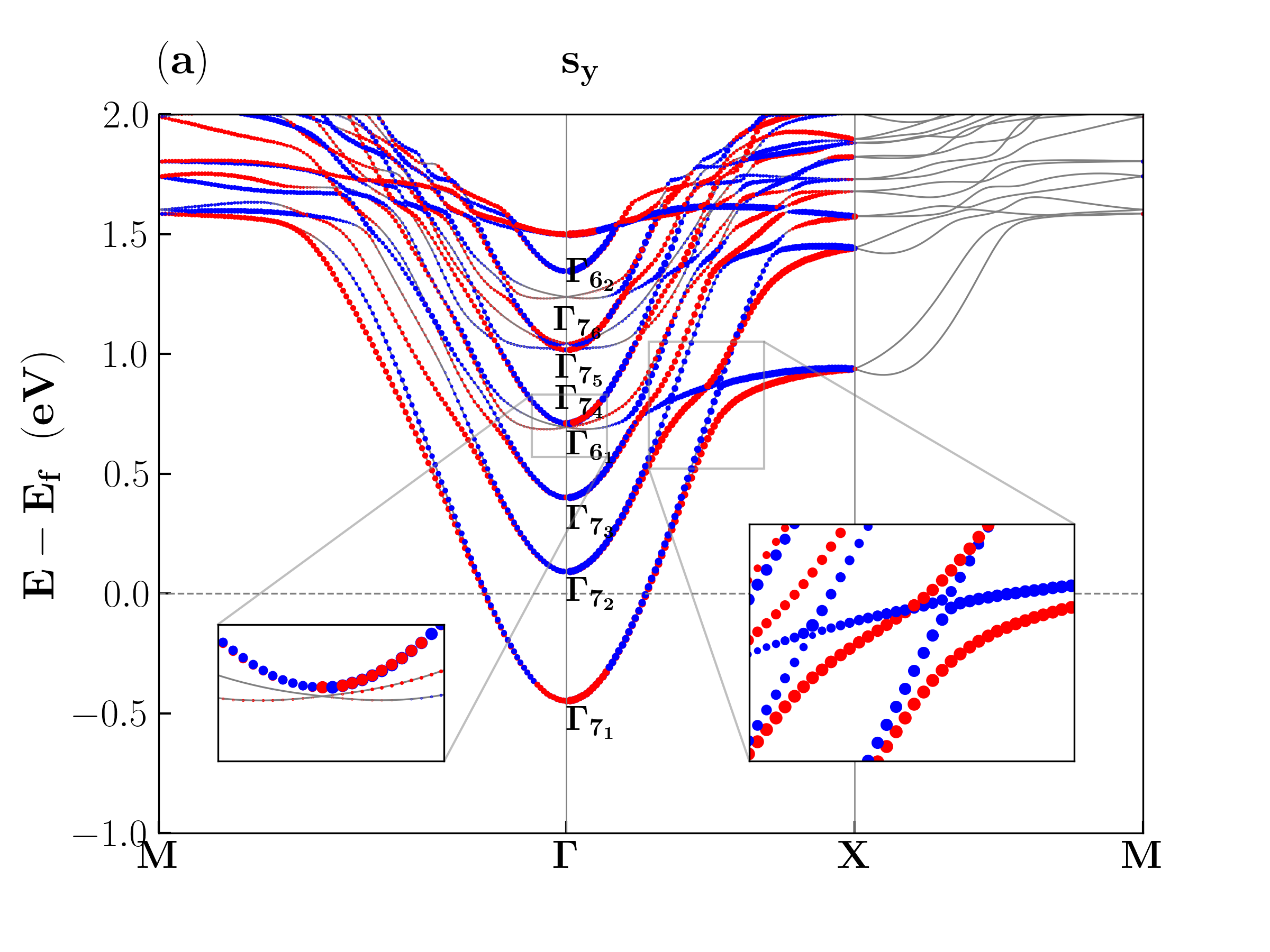}\hspace{-2.95cm}
		\includegraphics[width=0.5\textwidth,height=0.3\textheight,trim={1cm 1cm 6cm 0.85cm},clip=true]{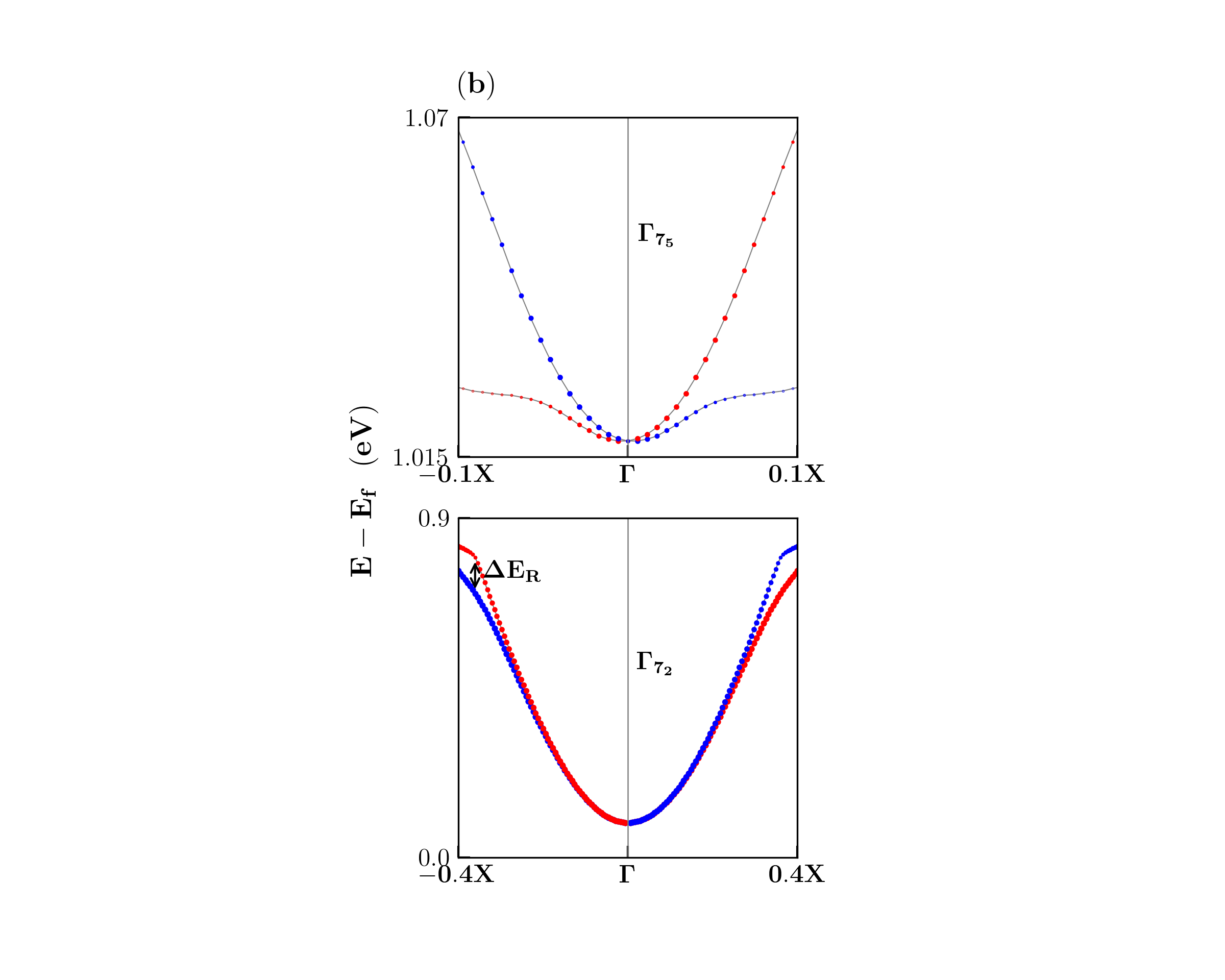}
		\caption{(a) Spin-orbit coupling (SOC) included Ta-5d band structure with the projection of electron spins along the M-$\Gamma$-X-M 
			direction, where non-degenerate components of Ta-$\rm{t_{2g}}$ bands are indicated by $\Gamma_{7}$ and $\Gamma_{6}$. 
			Red and blue colored circles represent positive and negative spin polarization along y direction. 
			The circle size represent the weight of the spin. M-O crossing is shown in the inset of (a).
			(b) Rashba spin splitting are shown for the $\Gamma_{7_2}$ and $\Gamma_{7_5}$ bands. 
			$\Delta{E_{R}}$ represents the maximum splitting. The direction of spin projections are  perpendicular to x-direction.}
		\label{Fig:ntype-GX-band}
	\end{figure}
\begin{figure}[htbp!]
		\includegraphics[width=0.5\textwidth,height=0.3\textheight,trim={1.3cm 0.5cm 1cm 0.55cm},clip=true]{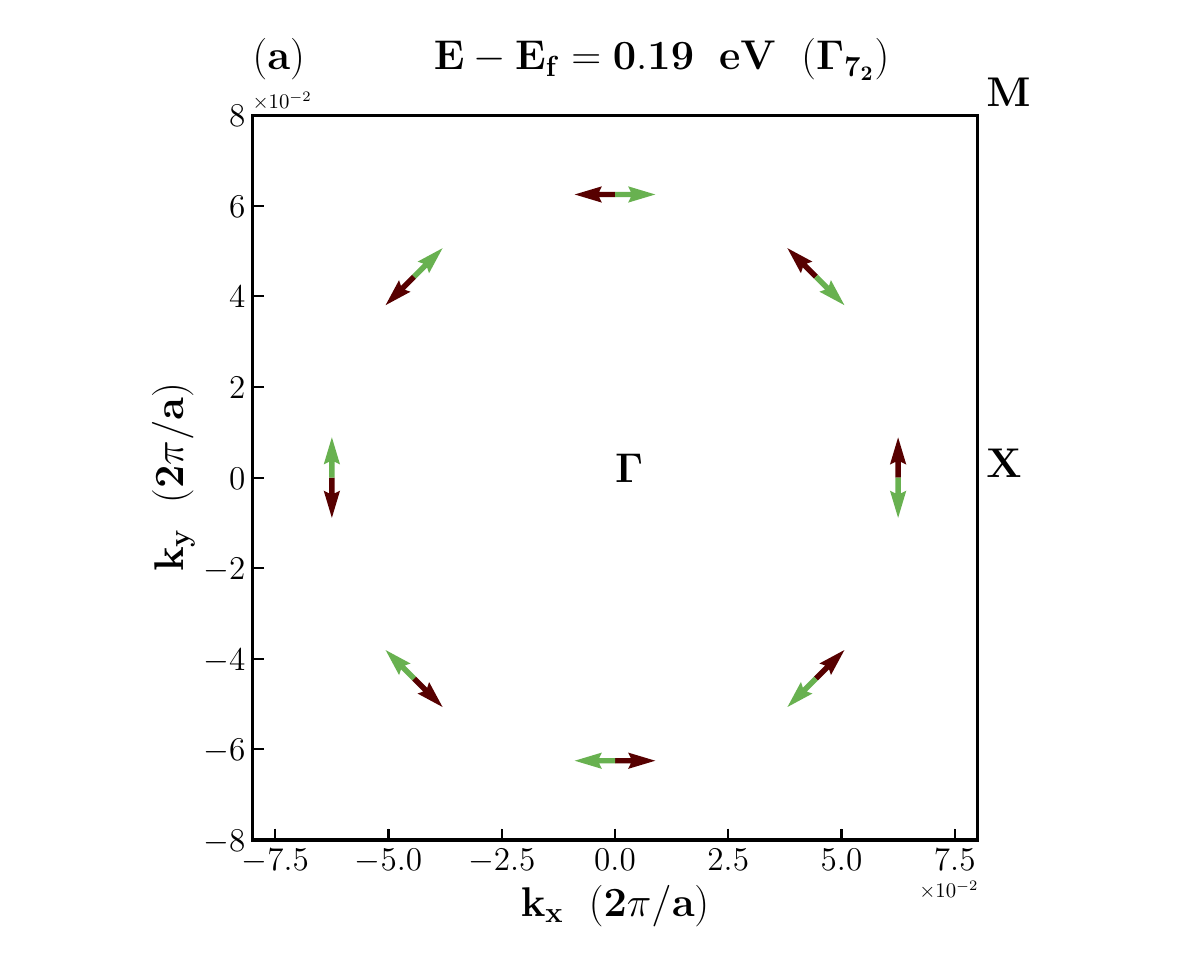}\hspace{-1cm}
		\includegraphics[width=0.5\textwidth,height=0.3\textheight,trim={0.55cm 0.5cm 1.3cm 0.55cm},clip=true]{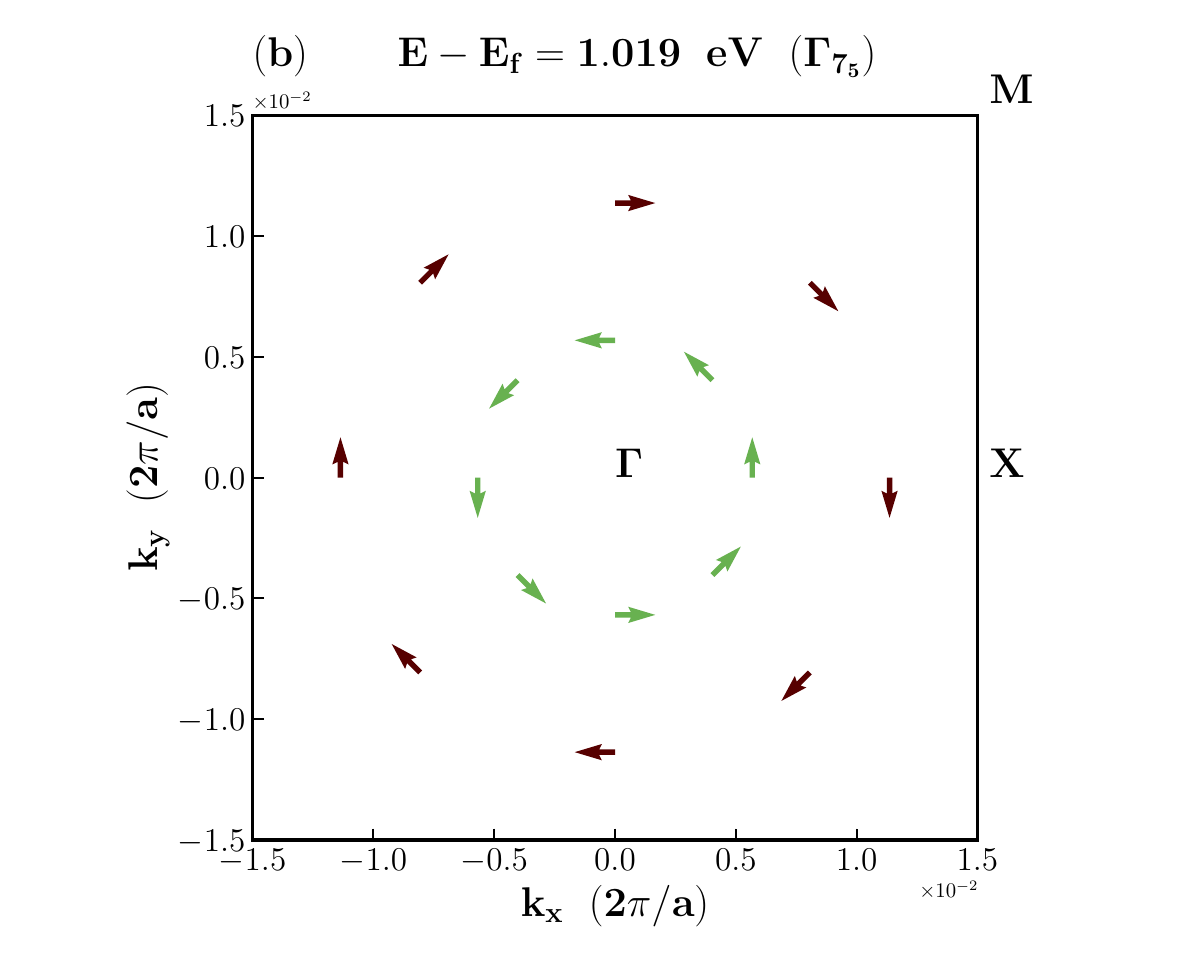}
		\caption{(a) Associated spin-texture of $\Gamma_{7_2}$ band in the 2D first Brillouin zone at iso-energy $\rm{E-E_{f}}=0.19~\textrm{eV}$, 
			(b) Spin-texture of $\Gamma_{7_5}$ at iso-energy $\rm{E-E_{f}}=1.019~\textrm{eV}$. Brown and green colors distinguish between the two splitted 
			bands with opposite spin orientation. The length of the arrows are in unit of $\hbar$.}
		\label{Fig:ntype-rso}
	\end{figure}

	Fig.~\ref{Fig:ntype-GX-band}(a) shows the band structure along $\rm{M}-\Gamma-\rm{X}-\rm{M}$ of the Type-I interface of LAO/KTO heterostructure in the presence of  SOC. 
	The spin polarizations are shown (solid red and blue circles) for the $\rm{s_y}$ component of the spin angular momentum.
	In Appendix~\ref{spr-I}, the spin polarization of $\rm{s_x}$ and $\rm{s_z}$ states are shown in Fig.~\ref{Fig:nt-split-band}(a) and (b).
	The spin-splitted Ta-5d orbitals are presented in the left panel. 
	According to the earlier report, the amount of RSO splitting is dependent on the M-O effect present
	in the system~\cite{soc-2013-sto,13-khalsa-MO,2014quasiparticle-MO,kong2021tunable}. In Ref.~\cite{kong2021tunable}, it is found that larger the M-O effect smaller is the RSO splitting. 
	To study the RSO effect, we have confined our study only along the $\Gamma-\rm{X}$ path.
	In the presence of SOC, all the Ta-5d bands break into the $\Gamma_{7}$ and $\Gamma_{6}$ levels. $\Gamma_{7}$ levels have orbital character of 
	$\rm{d_{xy}}$. In Fig.~\ref{Fig:ntype-GX-band}(a), we have presented all the $\Gamma_{7}$ levels and two $\Gamma_{6}$ levels. 
	$\Gamma_{7_1}$ level originates from the $\rm{d_{xy}}$ subband of subsrate layer (SL-I), and it has no spin-splitting.
	Systematic study of every $\Gamma_{7}$ level reveals that the spin-splitting gets stronger as we move away from the interface.  
	$\Gamma_{7_2}$ level has almost no splitting near k=0, whereas for larger k value (around 0.28~\AA) it has a splitting at
	the band crossing region, and it is predominantly cubic like with a RSO coupling strength $\alpha_{R3}=80~\textrm{eV\AA}^3$. 
	$\Gamma_{7_3}$ has a similar nature like $\Gamma_{7_2}$ level. The RSO interaction is cubic like with $\alpha_{R3}=0.98~\textrm{eV\AA}^3$ at
	$\rm{k_x} \approx 0.13~\textrm{\AA}$. In the $\Gamma_{7_5}$ and $\Gamma_{7_6}$ levels, for $k_x \lesssim 0.02 \textrm{\AA}$ the linear RSO 
	interaction strength is one order magnitude higher than the cubic type RSO with $\alpha_{R1} = 43~\textrm{meV\AA}$ and 
	$\alpha_{R1} =260~\textrm{meV\AA}$. 
	It is worthy to mention that in a recent study on $TaAs_{2}$ semimetal, the maximum spin splitting in Ta has been found to be 269 meV~\cite{wadge2022ta}. 
	In this article we have found that a larger splitting in Ta can be achieved by using oxide heterostructure of Ta-based material.
	$\Gamma_{7_4}$ level shows a spin flipping across the $\Gamma-\rm{X}$ path (shown in the left inset of Fig.~\ref{Fig:ntype-GX-band}(a)). However, there is no spin-splitting of the bands, and it shows 
	similar behavior to the $\Gamma_{7_1}$ level. The $\Gamma_{6}$ level, which is a complicated combination of $\rm{d_{yz}}$/$\rm{d_{xz}}$ and 
	O-2$\rm{p_{y}}$ subbands, shows a spin-slitting. However, this spin-splitting is not Rashba-like.
	We have estimated the RSO coupling strengths of each spin-splitted levels by fitting Eq.~\ref{fitting} to our
	DFT data, and the results are presented in Table~\ref{tab1}.
	\begin{table}[t]  
		\caption{The table represents the RSO coupling strength obtained by fitting Eqn.~\ref{fitting} with calculated DFT band dispersion data along $\Gamma-\rm{X}$.
			Corresponding plots are shown in supplemental material.}
		\centering  
		\begin{tabular}{c c c c c c} 
			\hline\hline   
			Orbital  & $\alpha_{R1}$ & $\alpha_{R3}$\\
			(Band Level)       	 &   (eV\AA)    &   (eV\AA$^3$) \\
			
			\hline\hline   
			$\rm{d_{xy}} ({\Gamma_{7_1}})$ & -- & -- \\
			$\rm{d_{xy}} ({\Gamma_{7_2}})$ & 0.8$\times10^{-3}$ & 0.80 \\
			$\rm{d_{xy}} ({\Gamma_{7_3}})$ & 2.4$\times10^{-3}$ & 0.98 \\
			$\rm{d_{xy}} ({\Gamma_{7_4}})$ & -- & -- \\
			$\rm{d_{xy}} ({\Gamma_{7_5}})$ & 0.04 & 39.14 \\
			$\rm{d_{xy}} ({\Gamma_{7_6}})$ & 0.26 & 113.42 \\
			\hline
		\end{tabular}  
		\label{tab1}
	\end{table}
	To study the nature of the spin-splitting we have plotted spin texture of the corresponding bands in the $\rm{k_x}-\rm{k_y}$ 
	plane at a particular energy-cut (light green arrows present the inner band spin orientations and brown arrows present the outer band
	spin orientations.)
	Fig.~\ref{Fig:ntype-rso}(a) shows the spin texture of $\Gamma_{7_2}$ band at iso-energy $\rm{E}=\rm{E_{f}}$+0.19 eV. 
	It is evident that at lower energy, around k=0, there is no spin-splitting of $\Gamma_{7_2}$ band. However, as pointed out 
	earlier, at about $\rm{E}=\rm{E_{f}}+0.69 \rm{eV}$, there is a lateral shift of up and down spin for inner and outer energy 
	bands (figure not shown). Fig.~\ref{Fig:ntype-rso}(b) shows the spin texture of $\Gamma_{7_5}$ level at $\rm{E}=\rm{E_{f}}$+1.019 eV.
	This spin texture reconfirms the linear RSO interaction in these bands up-to $\rm{k_x} \approx 0.02~\textrm{\AA}$. 
	\subsection{Type-II heterostructure without SOC}\label{t2-nso}
	\begin{figure}[htbp!]
		\includegraphics[height=0.3\textheight,width=0.5\textwidth,trim={0.5cm 1cm 0.3cm 0.85cm},clip=true]{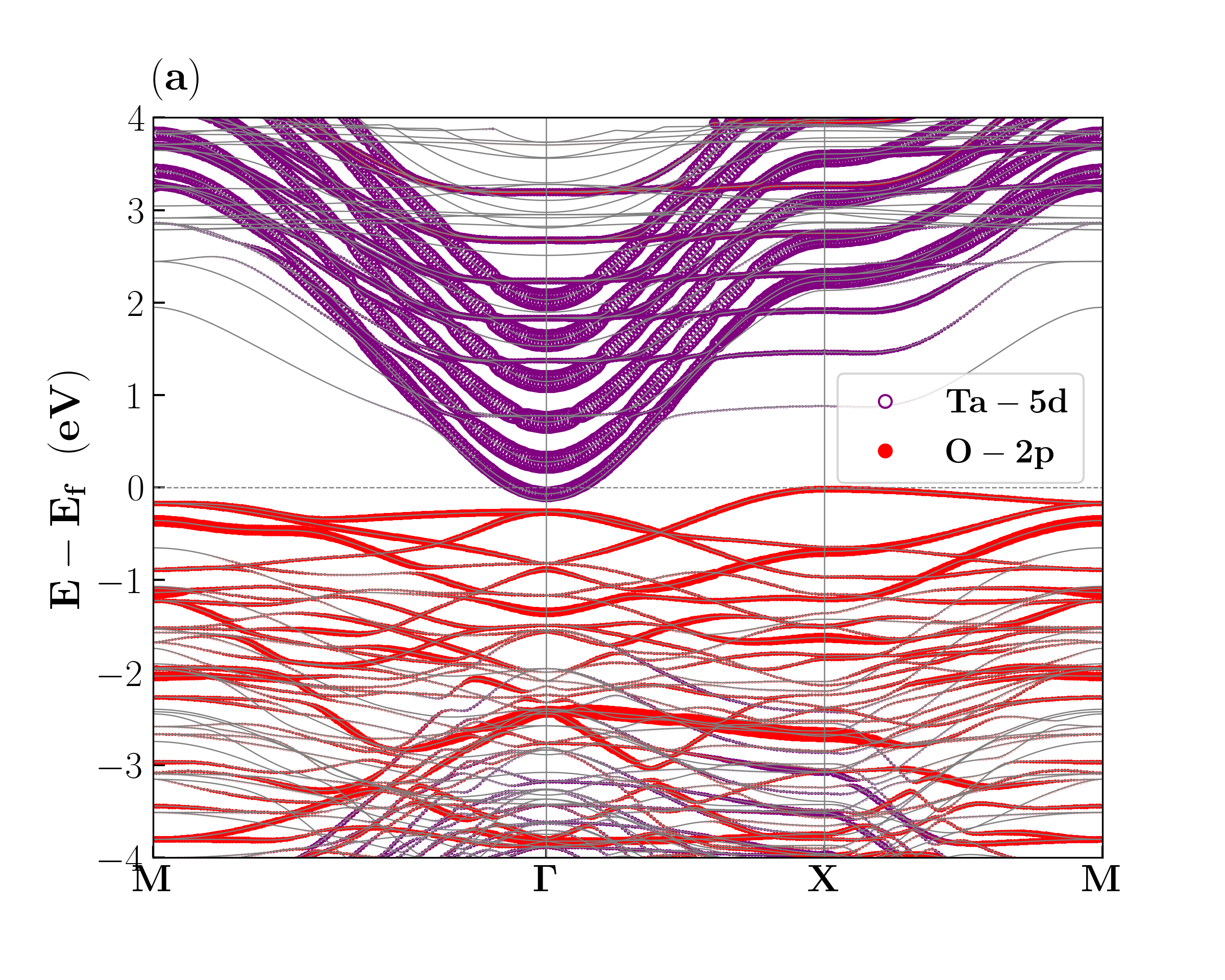}\hspace{-0.6cm}
		\includegraphics[height=0.3\textheight,width=0.5\textwidth,trim={0.5cm 1cm 0.3cm 0.85cm},clip=true]{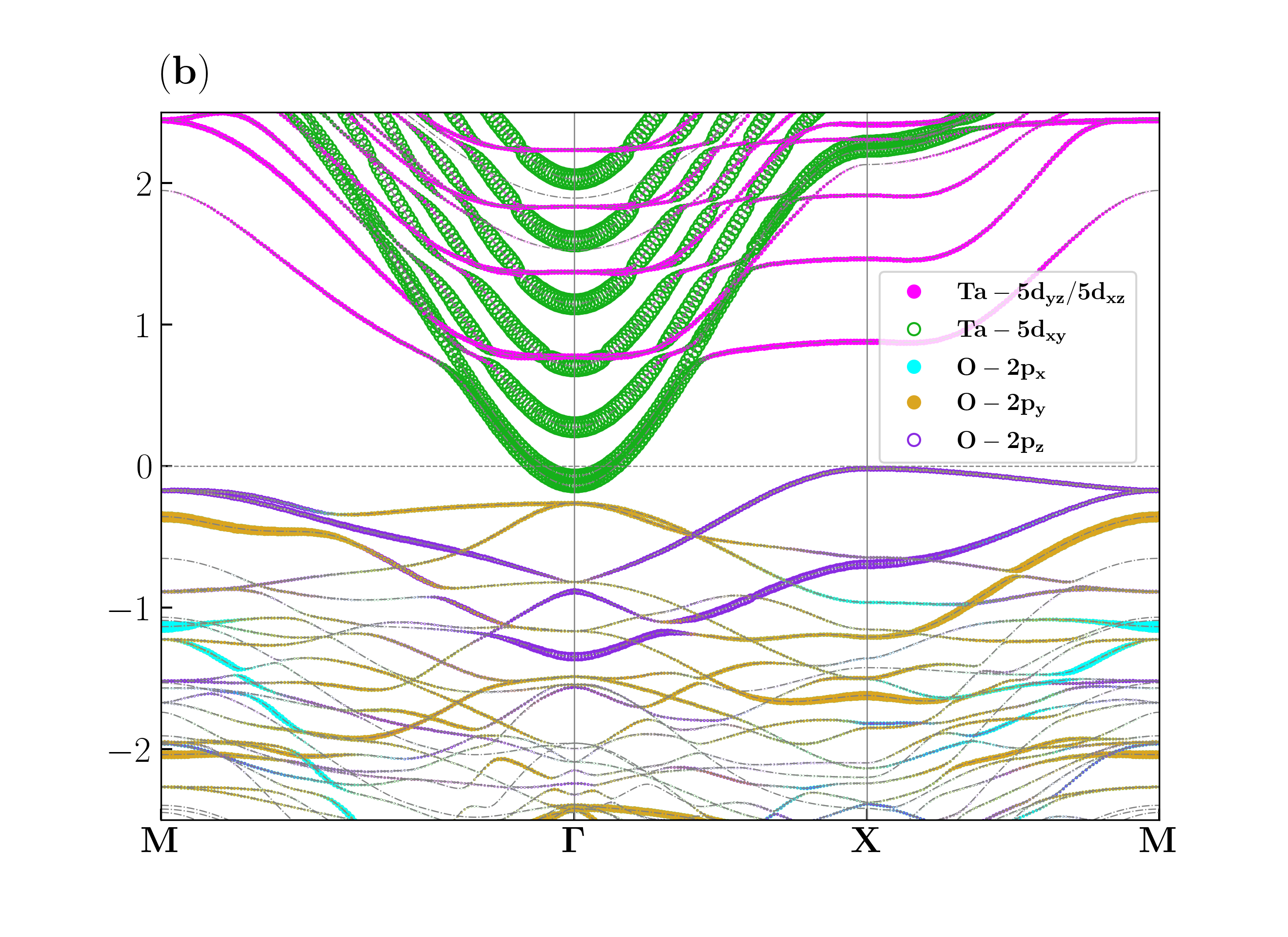}
		\includegraphics[height=0.67\textheight,width=0.55\textwidth,trim={0.3cm 1.4cm 1cm 3cm},clip=true]{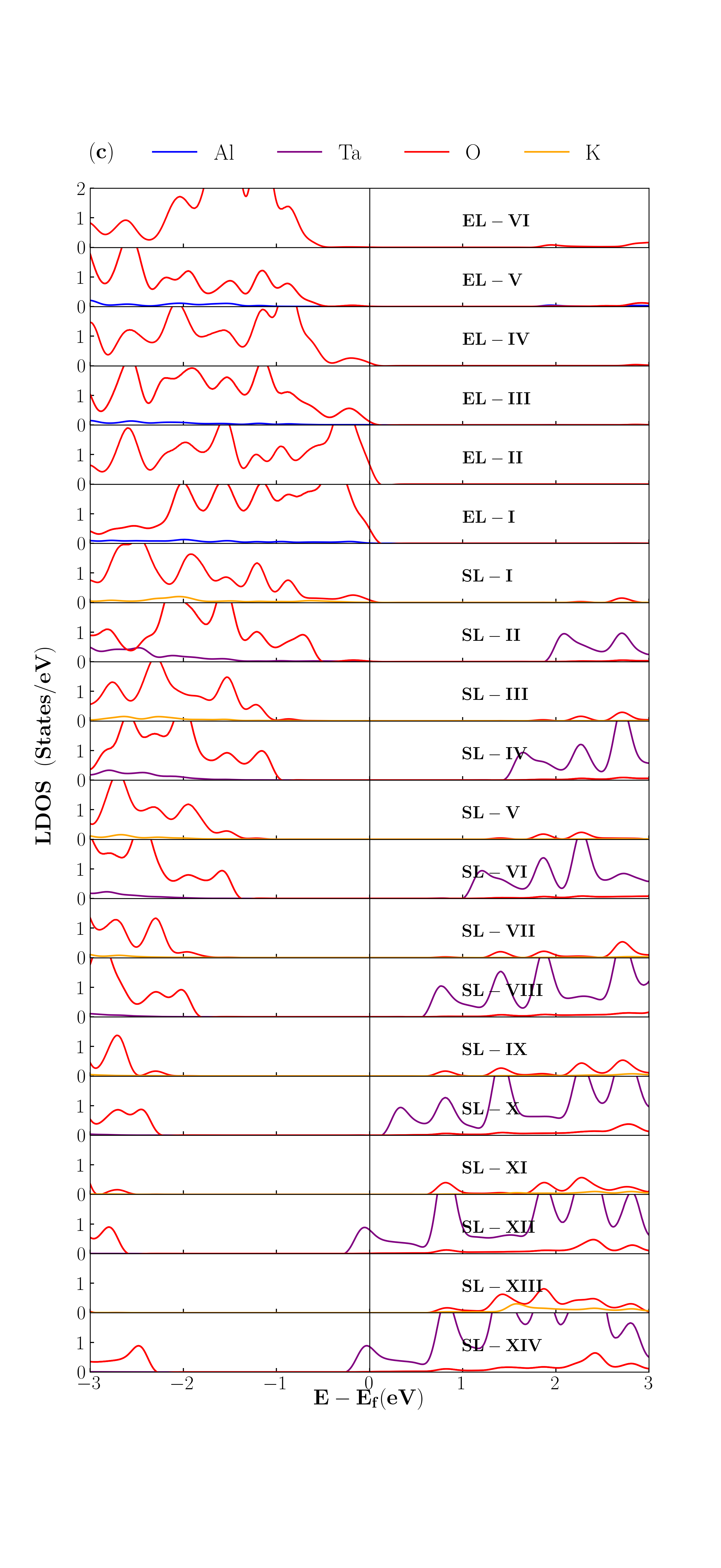}\vspace{-1cm}
		\caption{(a) Band structure of the Type-II heterostructure without SOC. (b) Zoomed-in display of the subbands coming from Ta-5d and O-2p orbitals, and (c) Layer-resolved density of states for the slab system.}
		\label{Fig:ptype-band}
	\end{figure}
	Fig.~\ref{Fig:ptype-band}(a) presents the band structure for Type-II system without SOC along $\rm{M}-\Gamma-\rm{X}-\rm{M}$ high symmetry path. 
	O-2p orbitals (marked with solid red circles) are originated from the polar interface created by $\text{KO}^{-}$ and ${\text{AlO}_\text{2}}^{-}$.
	Band structure shows that the conduction bands are consisted of Ta-5d subbands (marked by open purple circle) which arise from the $\text{KTaO}_\text{3}$ substrate. 
	The parabolic nature of conduction bands around $\Gamma$ indicates that there is a creation of quantum well.
	All the epitaxial and substrate layers are allowed to be relaxed in this heterostructure. 
	Coexistence of electron and hole gas have been observed at $\Gamma$ and $\rm{X}$ point respectively, 
	in contrast to the earlier reported result~\cite{Zhang-PCCP-Strain-18} for $\text{KTaO}_\text{3}$ thin film,
	where similar coexistence is reported at $\Gamma$ and $\rm{M}$ respectively.
	Furthermore, we have found that this qualitative feature remains unaltered in the case 
	when the last substrate layer is kept fixed. These results have been presented in the supplemental information.
	In Fig.~\ref{Fig:ptype-band}(b) we have presented the individual orbital contributions of Ta-5d and O-2p orbitals in the system.
	It is evident that similar to the Type-I heterostructure, the the degeneracy of the $\rm t_{2g}$ subband of Ta-5d orbital is lifted 
	in an identical manner. Moreover, there is a similar inter-orbital crossing of $\rm d_{xy}$ and $\rm d_{yz}$/$\rm d_{xz}$ bands 
	in this Type-II hetero-interface. In this oxygen-rich interface CFS also plays a crucial role in O-2p orbital splitting.
	After orbital separation, the valence bands become an admixture of $\rm{p_y}+\rm{p_x}$ orbitals originated from SL-I and $\rm{p_y}+\rm{p_z}$ orbitals, originated from EL-I and EL-II.
	LRDOS for $\text{KO}^{-}$/$\text{AlO}_\text{2}^{-}$ heterostructure reconfirms the formation of O-rich interface, which presented in 
	Fig.~\ref{Fig:ptype-band}(c). Since both the substrate and the epitaxial layers at the interface are made of negatively charged surfaces, 
	the occurrence of hole gas is expected. This can be seen from the valence band maxima (VBM) at X (0.5,0,0) (Fig.~\ref{Fig:pt-spin-XM} shown in Appendix~\ref{spr-II}) 
	which is primarily composed of O-$\rm{p_y}+\rm{p_z}$ orbitals.  
	The carrier density in this interface is found to be 3.35$\times$ 10$^{14}$ cm$^{-2}$.
	\subsection{Type-II heterostructure with SOC}\label{t2-rso}
	\begin{figure}[htbp!]
		\includegraphics[width=0.5\textwidth,height=0.3\textheight,trim={0.1cm 1cm 1cm 0.5cm},clip=true]{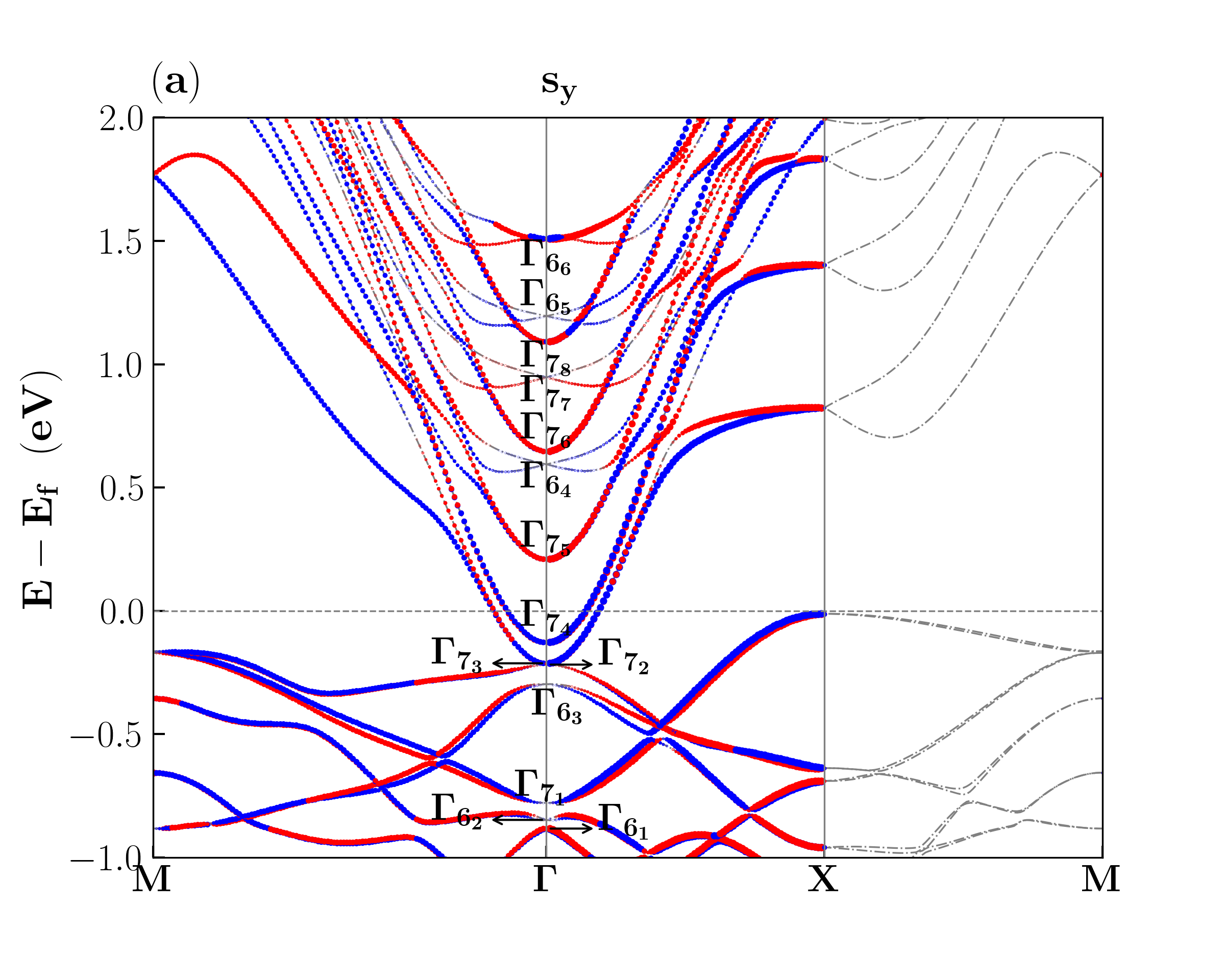}\hspace{-0.65cm}
		\includegraphics[width=0.5\textwidth,height=0.3\textheight,trim={0.1cm 1cm 1cm 0.5cm},clip=true]{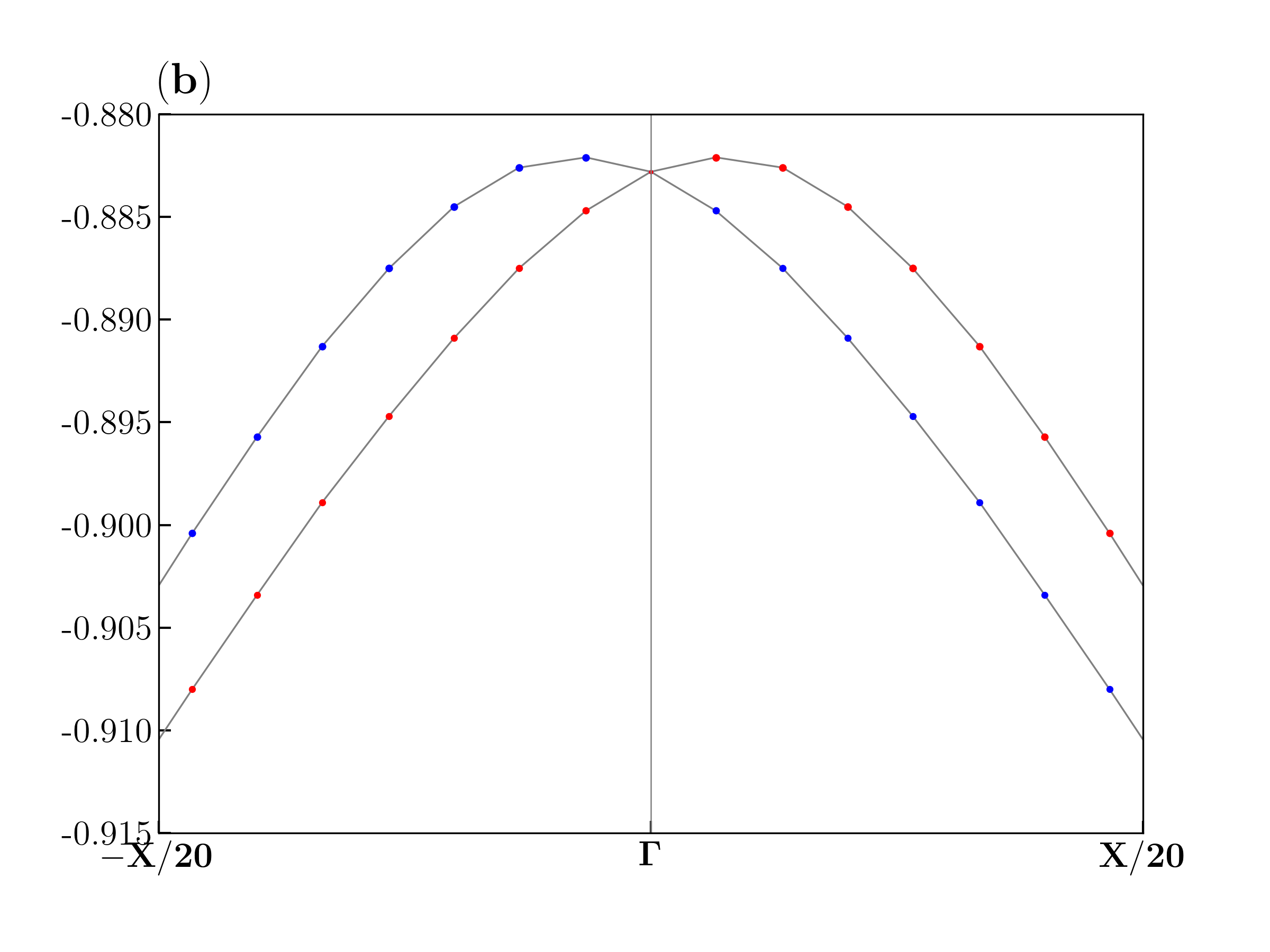}
		\caption{(a) SOC included band structure along $\rm{M}-\Gamma-\rm{X}-\rm{M}$ contributed by Ta-5d and O-2p orbitals. Red and blue colored circles represent up and down spin polarization along $s_y$. The circle size represent the weight of the spin. (b) Rashba spin-splitted band of $\Gamma_{6_1}$ level.}
		\label{Fig:ptype-spin-band}
	\end{figure}
	In this subsection, we have studied Type-II heterostructure in the presence of SOC, and the corresponding band structure is presented in the Fig.~\ref{Fig:ptype-spin-band}.
	The presence of SOC in Type-II systemÂ leads to the formation of the $\Gamma_{6}$ and $\Gamma_{7}$ levels.   
	Distinct $\Gamma_{6}$ and $\Gamma_{7}$ levels are indicated in the band structure by the numerals 1, 2, 3 etc.
	Unlike Type-I system, here both the $\Gamma_{6}$ and $\Gamma_{7}$ levels show RSO spin splitting.
	\begin{table}[t]  
		\caption{The table represents the RSO coupling strength for Type-II system obtained by fitting Eqn.~\ref{fitting} with calculated DFT band dispersion data.
			Corresponding plots are shown in supplemental material.} 
		\centering  
		\begin{tabular}{c c c c c c}   
			\hline\hline   
			Orbital  & $\alpha_{R1}$ & $\alpha_{R3}$\\
			(Band Level)&   (eV\AA)    &   (eV\AA$^3$) \\
			
			\hline\hline   
			O-{$\rm{2{p_y}+2{p_z}}$}($ {\Gamma_{6_1}}$) & 0.25 & -182.21 \\  
			Ta-5$\rm{d_{xy}}$ (${\Gamma_{7_3}}$) & 1.41$\times10^{-2}$ & -2.26 \\
			Ta-5$\rm{d_{xy}}$(${\Gamma_{7_4}}$) & 2.88$\times10^{-2}$ & -1.17$\times10^{-2}$ \\
			Ta-5$\rm{d_{xy}}$(${\Gamma_{7_5}}$) & 2.35$\times10^{-3}$ & 0.59 \\
			Ta-5$\rm{d_{xy}}$(${\Gamma_{7_6}}$) & 4.30$\times10^{-3}$ & 4.60$\times10^{-2}$ \\     
			Ta-5$\rm{d_{yz}/d_{zx}}$(${\Gamma_{6_6}}$) & 0.11 & -94.60  \\
			
			\hline
		\end{tabular}  
		\label{tab2r}
	\end{table} 
	This is an oxygen-rich heterojunction, and in contrast to the Type-I heterostructure, in this 
	system O-2p orbitals exhibit RSO-splitting in the valence band region.
	$\Gamma_{6_1}$ band is a hybridized state of O-2{$\rm{p_y}+\rm{p_z}$}. In the valence band region, only $\Gamma_{6_1}$ band shows Rashba-like spin splitting 
	with $\alpha_{R1}\approx250~\textrm{meV\AA}$ and $\alpha_{R3}\approx182.2~\textrm{eV\AA}^3$.
	Suprisingly, we have observed almost a three-fold increase in $\alpha_{R1}$ and more than ten-fold 
	decrease in $\alpha_{R3}$ values when the final substrate layer is kept fixed. These results have 
	been presented in the supplemental information.
	The spin-splitted $\Gamma_{6_1}$ band has been presented in Fig.~\ref{Fig:ptype-spin-band}(b). 
	$\Gamma_{7_3}$, $\Gamma_{7_4}$, $\Gamma_{7_5}$, $\Gamma_{7_6}$ are originated from Ta-5$\rm{d_{xy}}$ levels of different substrate layers.
	The RSO spin splitting of these levels are predominantly cubic in nature. 
	In $\Gamma_{6_2}$, $\Gamma_{6_3}$, $\Gamma_{6_4}$, $\Gamma_{6_5}$ and $\Gamma_{7_1}$, $\Gamma_{7_2}$, $\Gamma_{7_7}$, $\Gamma_{7_8}$ (which are not shown in Table~\ref{tab2r}) there a continuous flipping 
	between the two spin channels in the immediate neigbourhood of the $\Gamma$ point, and hence the 
	observed spin-splitting in these levels are not categorized as RSO spin splitting.
	$\Gamma_{6_6}$ level is originated from Ta-5$\rm{d_{yz}/d_{zx}}$ band and it has predominantly linear RSO coupling strength of $\alpha_{R1}\approx110~\textrm{meV\AA}$. 
	The spin channel in this band survives upto $\frac{X}{20}$ away from $k=0$. The linear and cubic Rashba coupling strengths for all the spin-splitted levels are presented in the Table~\ref{tab2r}.
	\begin{figure}[htbp!]
		\includegraphics[width=0.6\textwidth,trim={2.9cm 2.8cm 9.5cm 3.0cm},clip=true]{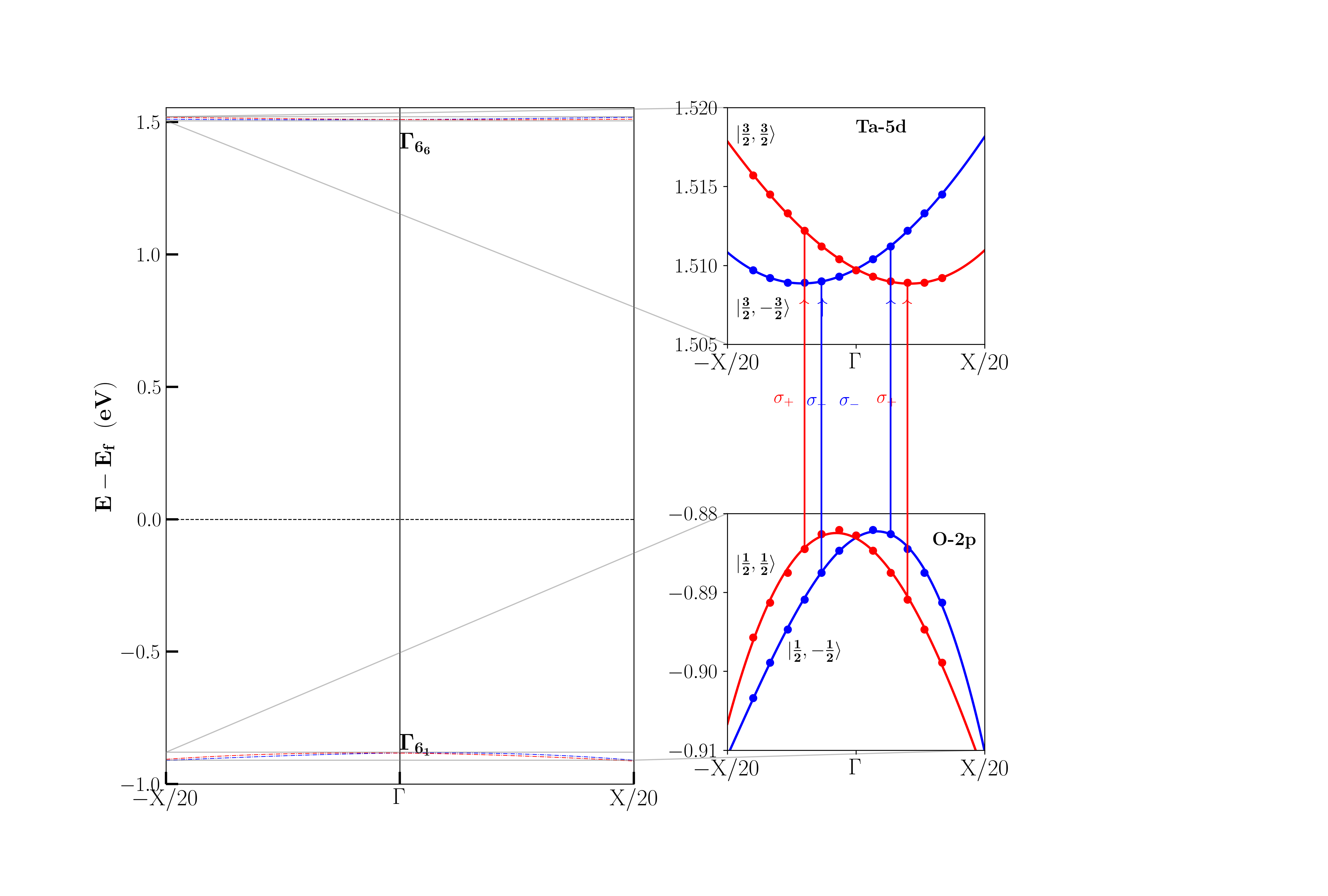}
		\caption{Schematic representation of circularly polarized photogalvanic effect for Type-II system. $\sigma_{+}$ (red) and $\sigma_{-}$ (blue) are the right handed and left handed polarized light. Red and blue arrows show the possible transitions.}
		\label{cpger}
	\end{figure}
	
	Interestingly, the Type-II system is found to be an ideal system to explore CPGE because it is a member of the $\rm{C_{4v}}$ point group and possesses a large k-linear RSO splitting with electron spin polarization along $\rm{k_x}$. Fig.~\ref{cpger} is a schematic representation of the possible electronic arrangement for producing circularly polarized photocurrent in the system which may provide a likely explanation for the recent experimental observation of CPGE in the LAO/KTO heterostructure~\cite{sto-lao-cpge}.
	In this system, the required optical transitions for CPGE take place between ${\Gamma_{6_1}}$ and 
	$ {\Gamma_{6_6}}$ levels. For the right-handed polarized light ($\sigma_{+}$), the interband transition is possible only for $\ket{j,m_j}$=$\ket{{\frac{1}{2}},{\frac{1}{2}}}\rightarrow\ket{{\frac{3}{2}},{\frac{3}{2}}}$ and similarly for left handed polarized light ($\sigma_{-}$) the only possible transition is $\ket{j,m_j}$=$\ket{{\frac{1}{2}},{-\frac{1}{2}}}\rightarrow\ket{{\frac{3}{2}},{-\frac{3}{2}}}$.  
	${\Gamma_{6_1}}$ and $ {\Gamma_{6_6}}$ levels are originated from the O-2p and Ta-5d bands of the EL-I and SL-IV, respectively.
	Surprisingly, this qualitative feature remains unaltered for the same system with fixed substrate layer. However, the transition route of the photocurrent is different for the later case. The results for Type-II
	system with fixed substrate layer have been presented elaborately in the supplemental material.
	\section{Conclusion}\label{Conlusion}
	In conclusion, $\text{TaO}_\text{2}^{+}$/$\text{LaO}^{+}$ and $\text{KO}_\text{2}^{-}$/$\text{AlO}_\text{2}^{-}$, both of which are referred 
	to as Type-I and Type-II stoichiometric interfaces, has been individually explored in this work. 
	We have investigated the band dispersion characteristics along the $\Gamma-\rm{X}$ path for both the systems without and with SOC.
	We have found that, Ta-5$\rm{d_{xy}}$ bands can produce the strongest k-linear RSO for Type-I systems with a strength 
	up to 260~$\textrm{meV\AA}$, while the $\rm{k^3}$-RSO coupling can attain a value up-to 113.42~$\textrm{eV\AA}^3$.
	In this Type-I heterostructure, no RSO splitting is observed in Ta-5$\rm{d_{yz}/d_{zx}}$ bands. 
	However, Type-II system predominantly produce k-linear RSO in both the Ta-5d and O-2p orbitals. The maximum linear RSO coupling strength for this system 
	is estimated to be 250~$\textrm{meV\AA}$ in O-2p band and 110~$\textrm{meV\AA}$ in Ta-5d band. Furthrmore,
	we have discovered a possible optical transition path in this Type-II system that may give rise to the
	circular photogalvanic effect, which is still quite rare in oxide heterostructures.
	\section{Acknowledgements}
	The authors would like to thank the computational facility provided by the National Institute of Technology, Rourkela. SB would like to thank Sri Sandipan Dasgupta, 
	Bhaba Atomic Research Centre, VECC, Kolkata, India, for very useful discussions. 
	\newpage
	%
	\clearpage
	\appendix
	\section{Spin polarized band-structures of Type-I}\label{spr-I}
	\begin{figure}[htbp!]
		\includegraphics[width=0.5\textwidth,height=0.3\textheight,trim={0.45cm 1cm 1cm 0cm},clip=true]{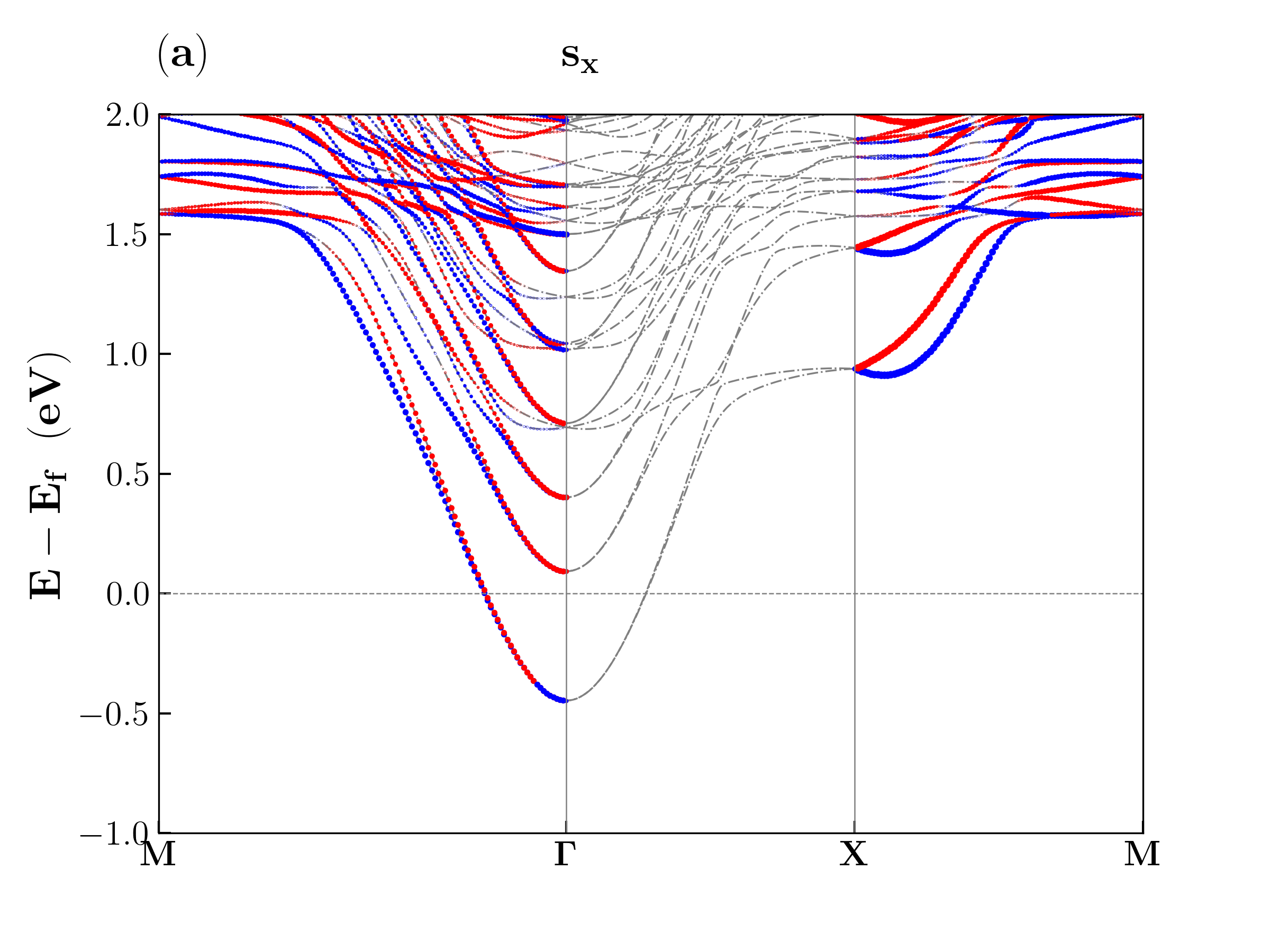}\hspace{-0.4cm}
		\includegraphics[width=0.5\textwidth,height=0.3\textheight,trim={1cm 1cm 1cm 0cm},clip=true]{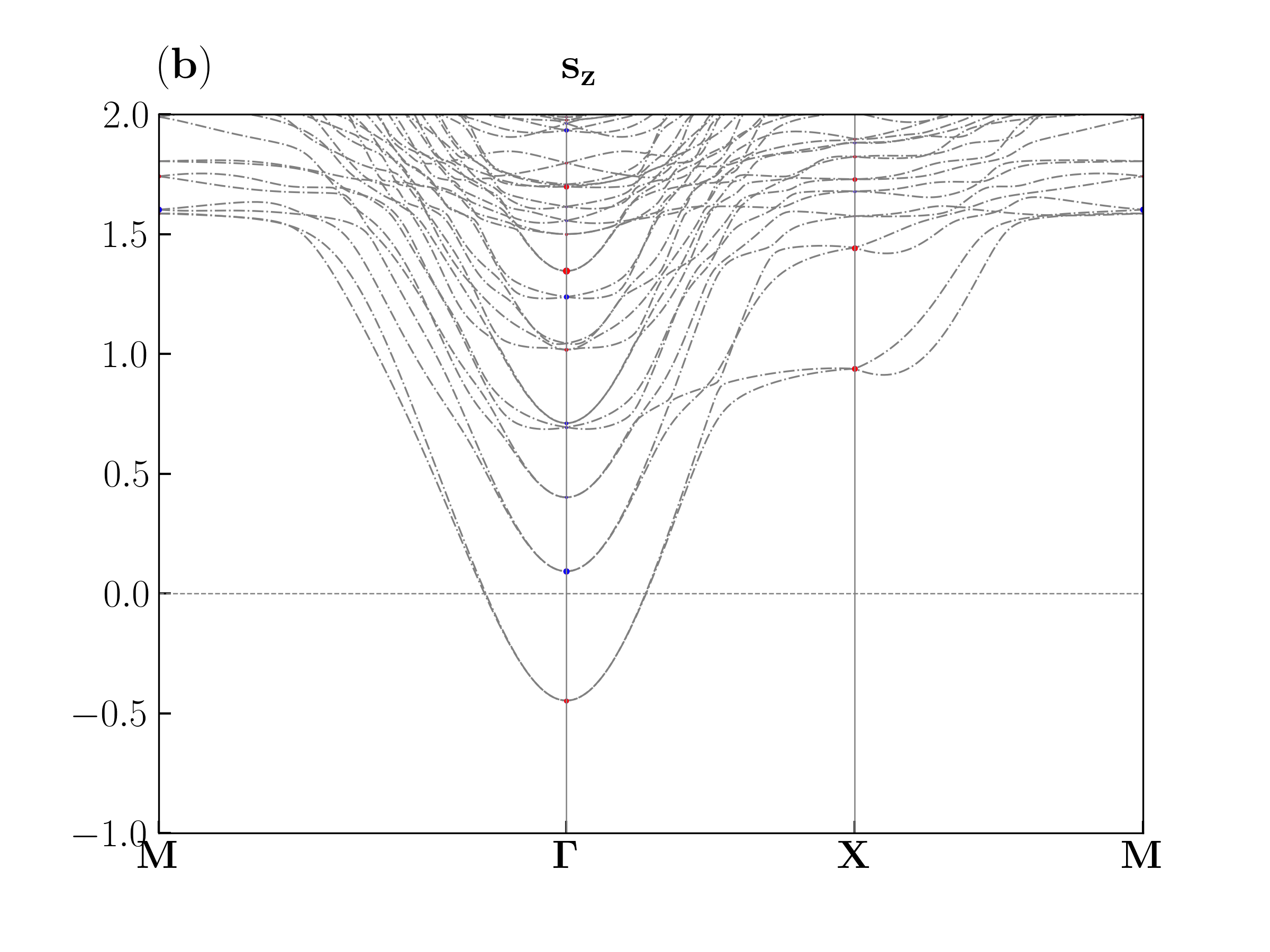} 
		\caption{(a) and (b) show SOC included band structures of the Type-I heterostructure with the projection of electron spins to $s_x$ and $s_z$, respectively. 
			Red and blue dots denote up and down components of spin projections, respectively. 
			The size of the dot shows the relative amplitude of the corresponding spin component.}
		\label{Fig:nt-split-band}
	\end{figure}
	Fig.~\ref{Fig:nt-split-band}(a) Fig.~\ref{Fig:nt-split-band}(b) present the band structure of Type-I system with the spin polarization along x and z direction.
	Along z direction there is no spin polarization present in the band dispersion. Along $s_x$, there is spin polarization, although we have not studied this system in detail.
	\section{Spin polarized band-structures of Type-II}\label{spr-II}
	\begin{figure}[htbp!]
		\includegraphics[width=0.5\textwidth,height=0.3\textheight,trim={0.3cm 1cm 1cm 0cm},clip=true]{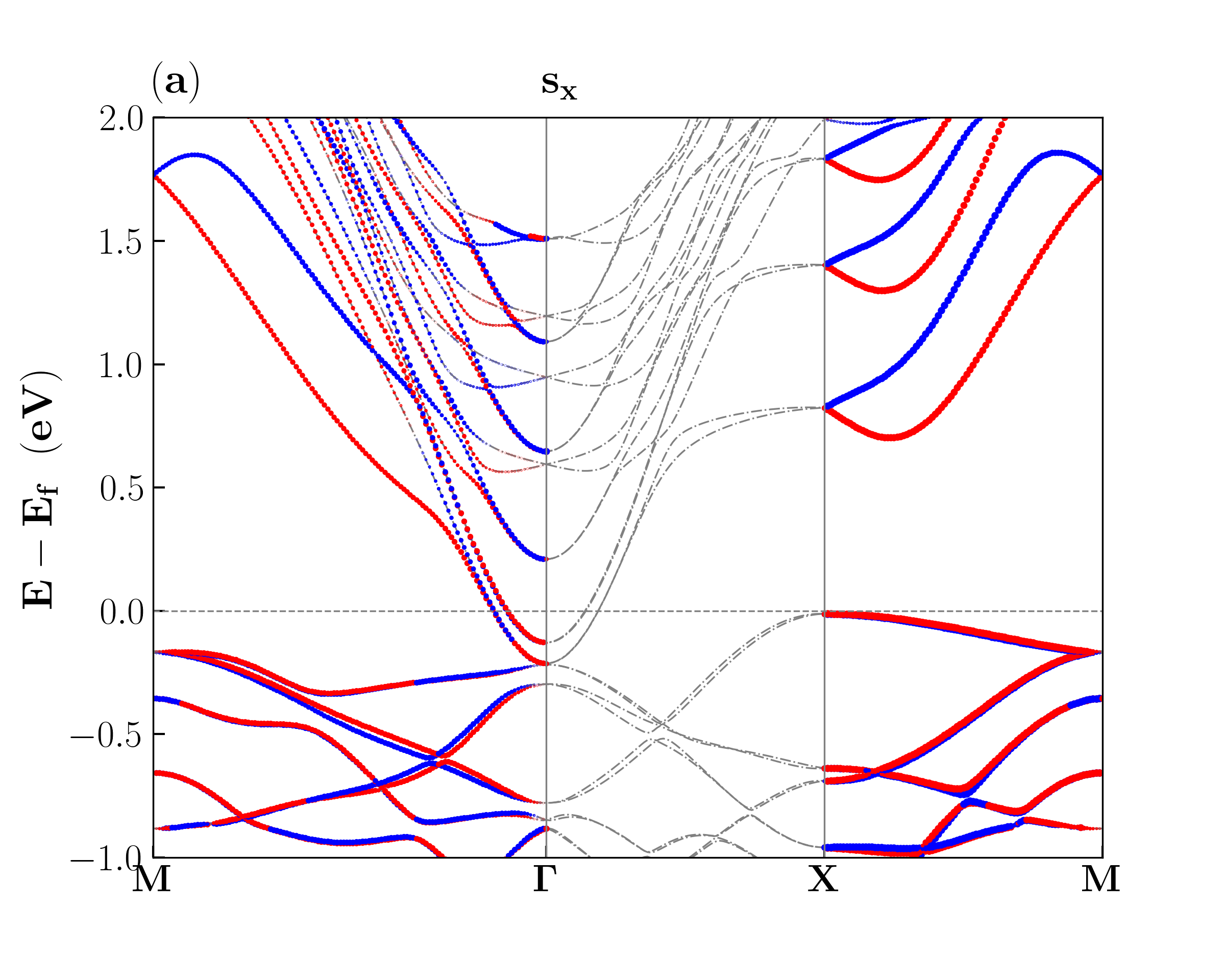}\hspace{-0.4cm}
		\includegraphics[width=0.5\textwidth,height=0.3\textheight,trim={1cm 1cm 1cm 0cm},clip=true]{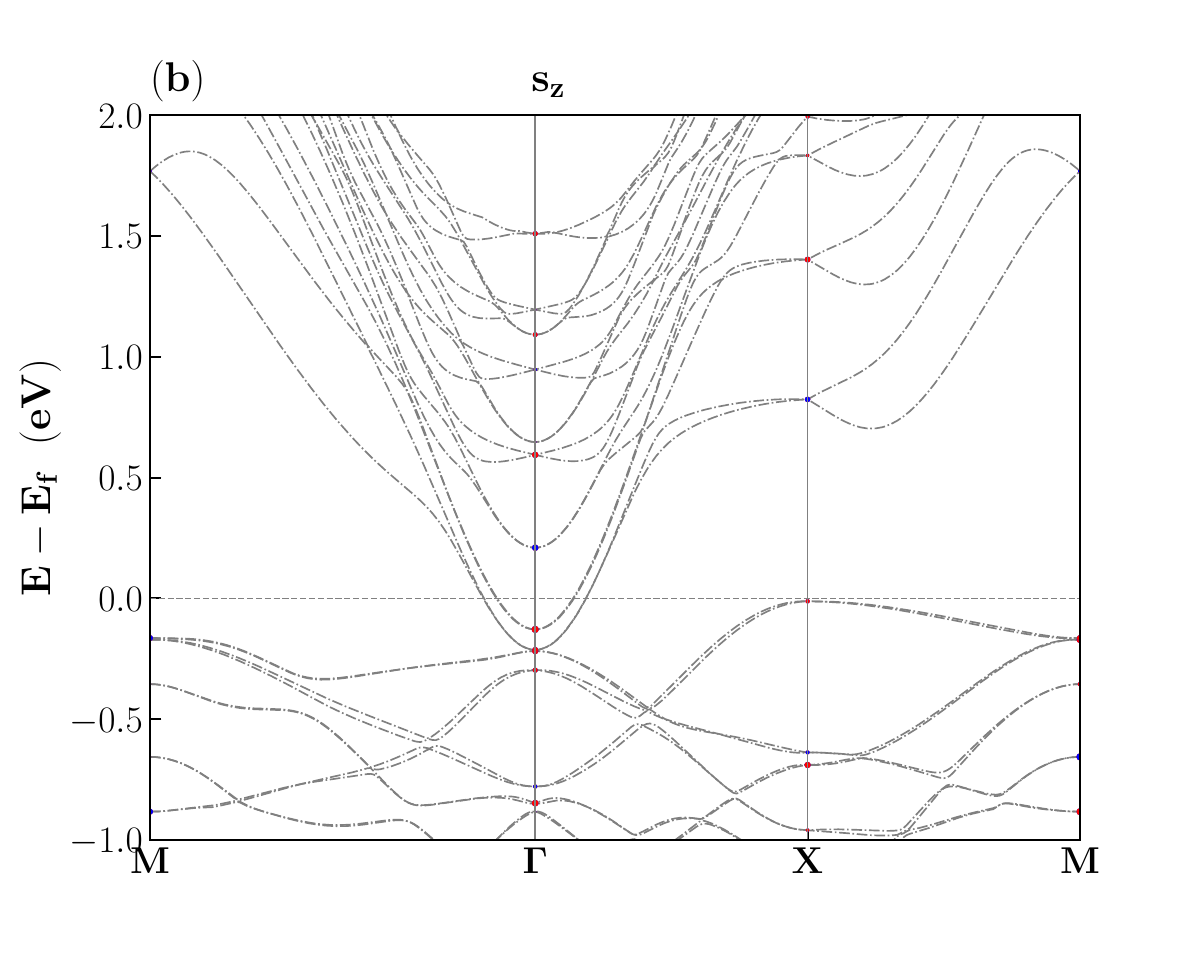} 
		\caption{(a) and (b) show SOC included band structures of the Type-II heterostructure with the projection of electron spins to $s_x$ and $s_z$, respectively. 
			Red and blue dots denote up and down components of spin projections, respectively. 
			The size of the dot shows the relative amplitude of the corresponding spin component.}
		\label{Fig:pt-split-band}
	\end{figure}
	Fig.~\ref{Fig:pt-split-band}(a) Fig.~\ref{Fig:pt-split-band}(b) show the Type-II system's band structure with spin polarization in the x and z directions.
	The band dispersion exhibits no spin polarization in the direction of z. Spin polarization exists along $s_x$, however we have not thoroughly examined this system.
	\begin{figure}[htbp!]
		\includegraphics[width=0.5\textwidth,height=0.3\textheight,,trim={1cm 0.5cm 1cm 0cm},clip=true]{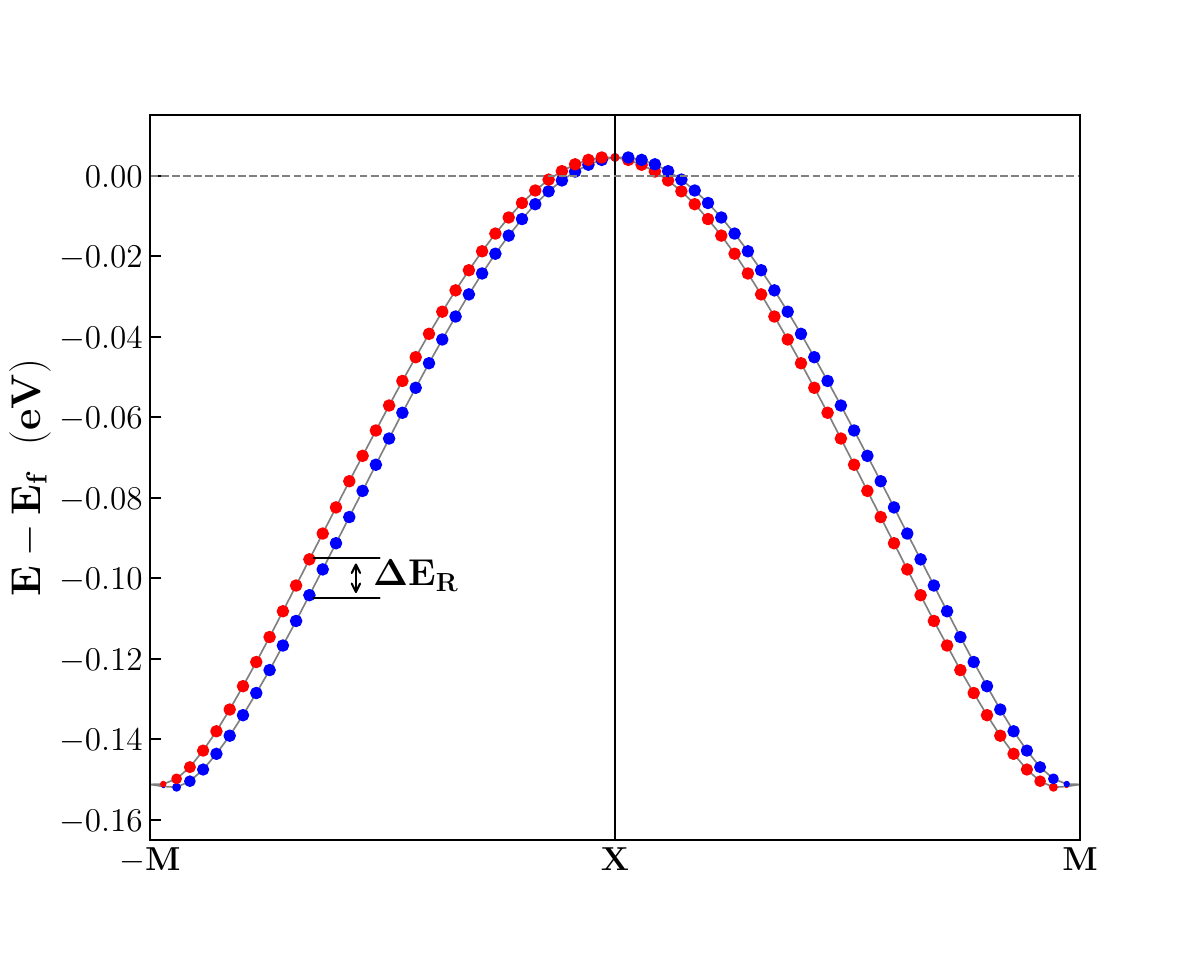}
		\caption{(a) SOC included band structures of the Type-II heterostructure along M-X-M with the projection of electron spins $s_y$. It represents that
			the VBM of Type-II system is appeared at the X (0.5,0,0) HS point.} 
		\label{Fig:pt-spin-XM}
	\end{figure}
	At the HS point X (0.5,0,0), the valence band maxima has been seen. The RSO splitting along the M-X-M HS path is cubic-like. This paper does not contain a full discussion of this observation since we  have restricted our analysis to the $\Gamma-\rm{X}$ path.
		\section{Fitting equation obtained from Hamiltonian}
	\begin{equation}
		\Delta_{R} = \epsilon_{1} - \epsilon_{2} = 2 \alpha_{R1} k_{x} + 2 \alpha_{R3} k_{x}^{3}
		\label{fitting}
	\end{equation}
	\subsection{Fitting for RSO parameter in Type-I heterostructure}
	\begin{figure}[htbp!]
		\centering
		\includegraphics[width=0.5\textwidth]{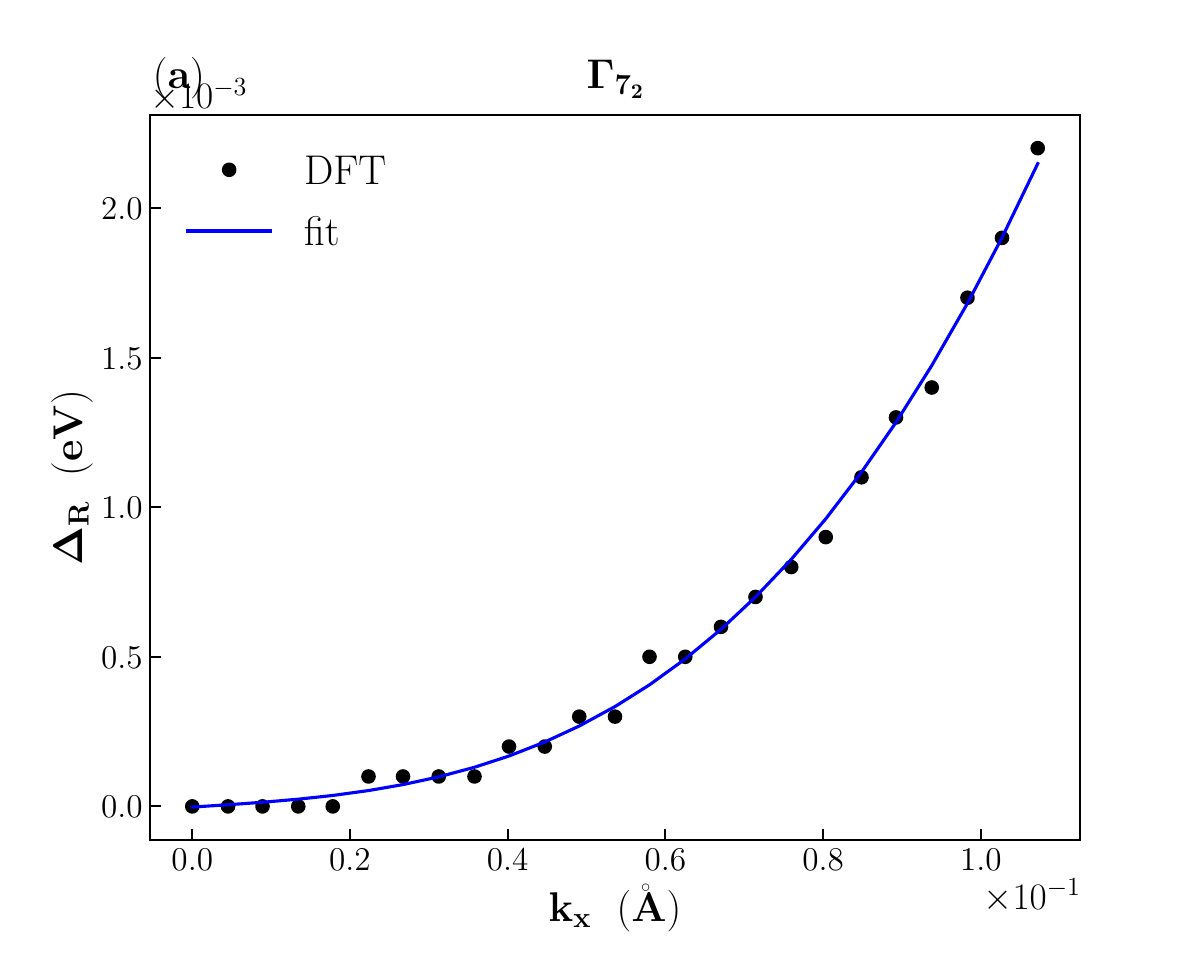}\hspace{-0.5cm}
		\includegraphics[width=0.5\textwidth]{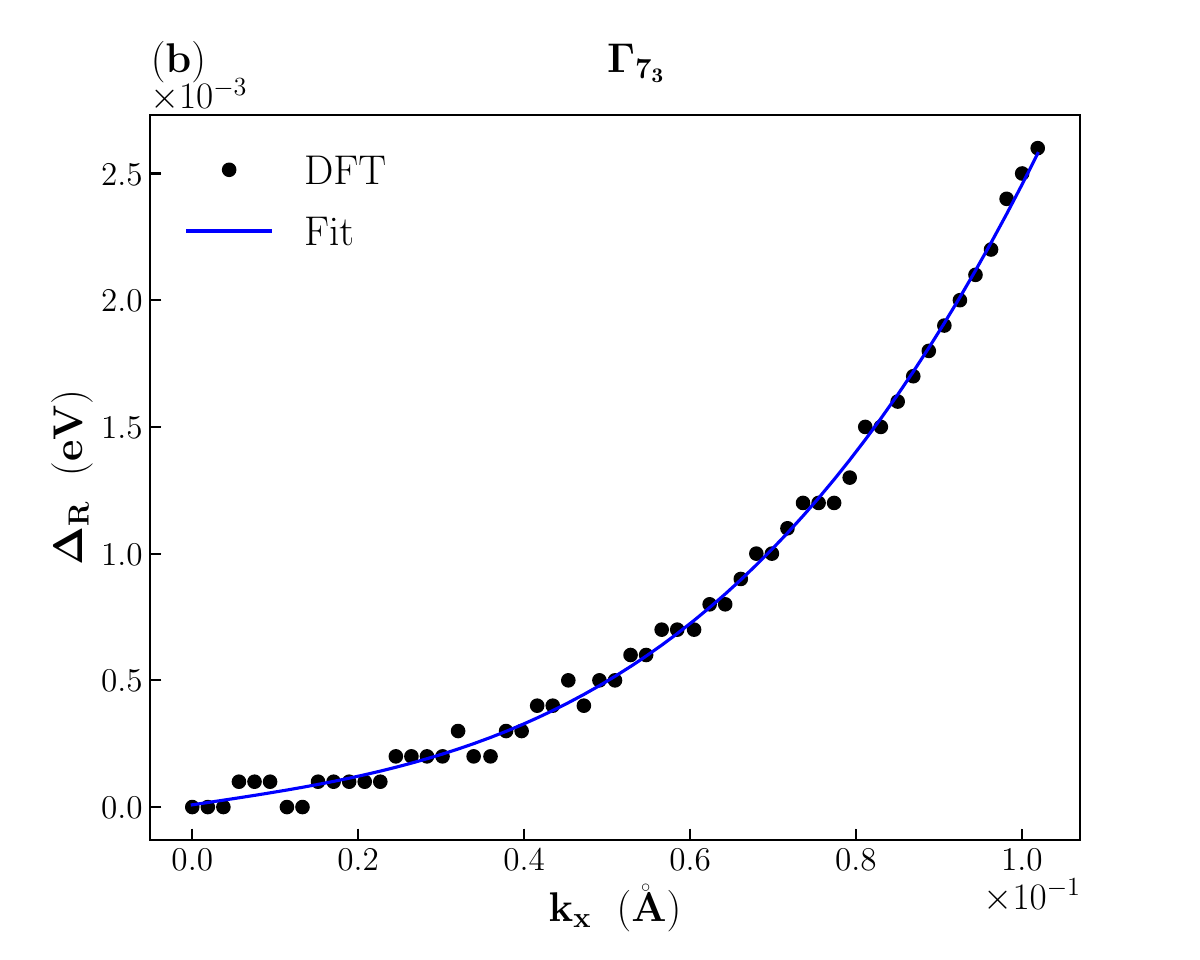}
		\includegraphics[width=0.5\textwidth]{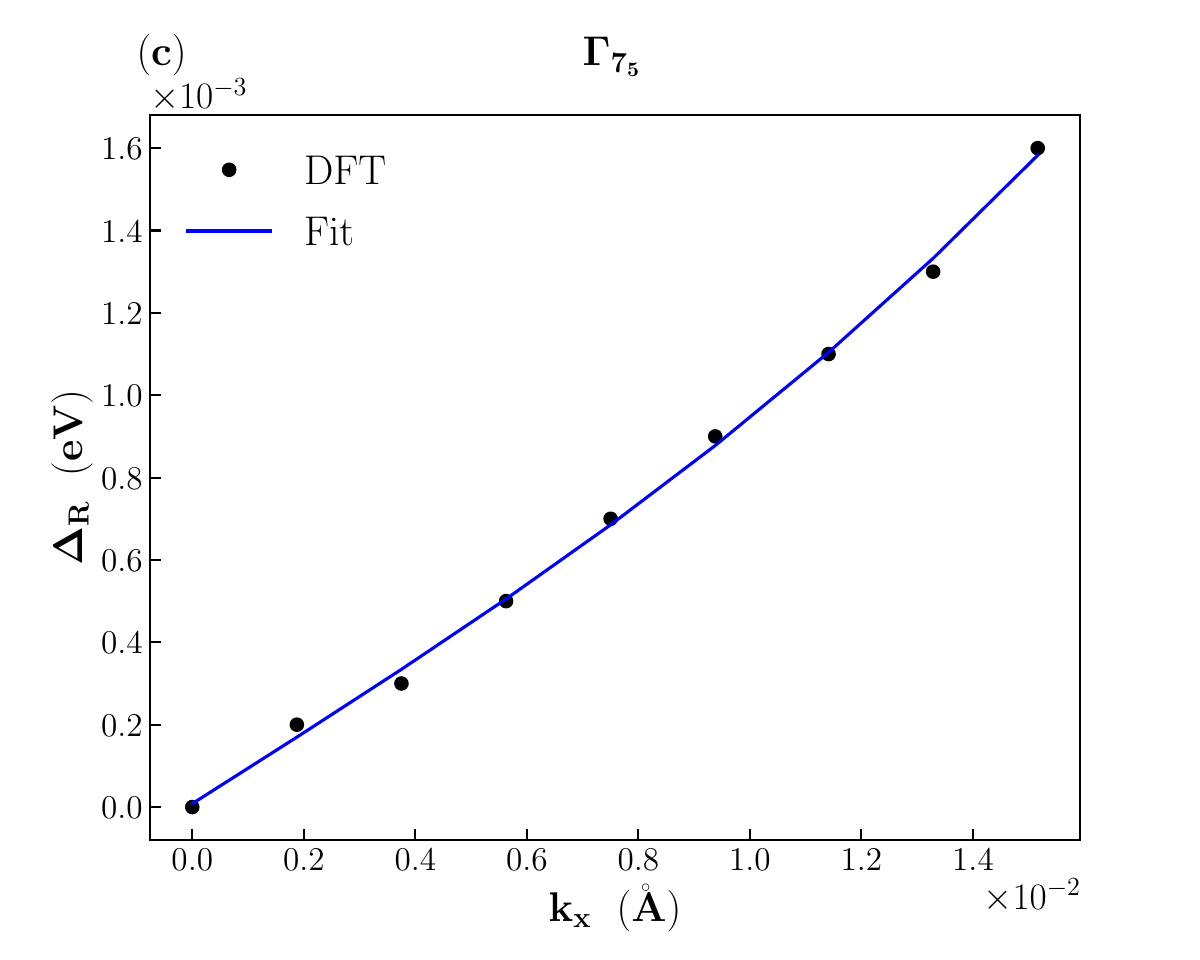}\hspace{-0.5cm}
		\includegraphics[width=0.5\textwidth]{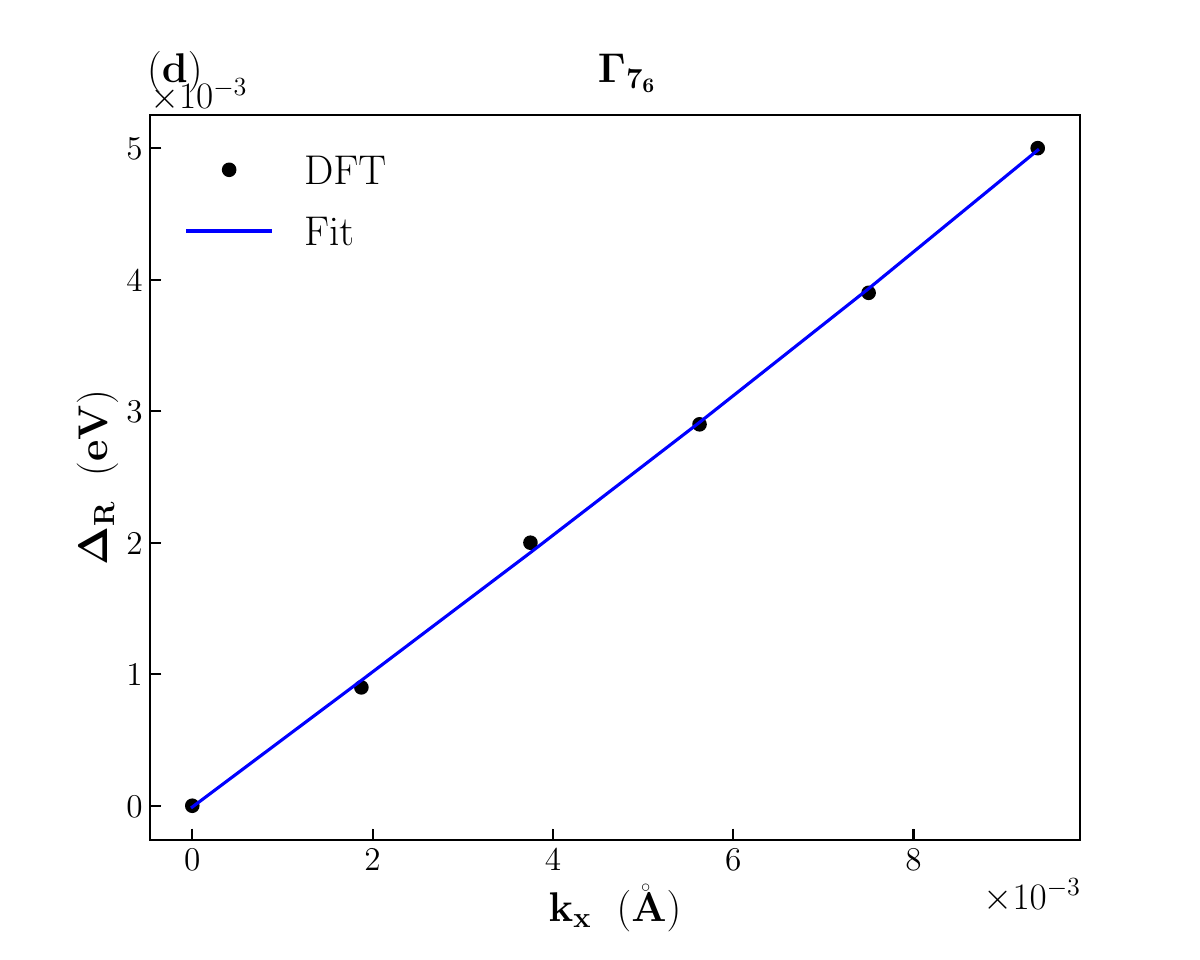}
		\caption{Hamiltonian fitting by using eqn.~\ref{fitting} for
			(a) $\Gamma_{7_2}$
			(b) $\Gamma_{7_3}$ 
			(c) $\Gamma_{7_5}$
			(d) $\Gamma_{7_6}$  bands}
		\label{Fig:ntype-fit-sp}
	\end{figure}
	\begin{table}[htbp!]  
		\caption{The table represents the RSO coupling strength obtained by fitting Eqn.~\ref{fitting} with calculated DFT band dispersion data along $\Gamma-\rm{X}$.
			Corresponding plots are presented in Fig.~\ref{Fig:ntype-fit-sp}.}
		\centering 
		\begin{tabular}{c c c c c c} 
			\hline\hline   
			Orbital  & $\alpha_{R1}$ & $\alpha_{R3}$\\
			(Band Level)       	 &   (eV\AA)    &   (eV\AA$^3$) \\
			
			\hline\hline   
			$\rm{d_{xy}} ({\Gamma_{7_1}})$ & -- & -- \\
			$\rm{d_{xy}} ({\Gamma_{7_2}})$ & 0.8$\times10^{-3}$ & 0.80 \\
			$\rm{d_{xy}} ({\Gamma_{7_3}})$ & 2.4$\times10^{-3}$ & 0.98 \\
			$\rm{d_{xy}} ({\Gamma_{7_4}})$ & -- & -- \\
			$\rm{d_{xy}} ({\Gamma_{7_5}})$ & 0.04 & 39.14 \\
			$\rm{d_{xy}} ({\Gamma_{7_6}})$ & 0.26 & 113.42 \\
			\hline
		\end{tabular}
		\label{tab1}
	\end{table}
	\newpage
	\subsection{Fitting for RSO parameter in Type-II heterostructure without any fixed substrate layer}
	\begin{figure}[htbp!]
		\centering
		\includegraphics[width=0.5\textwidth]{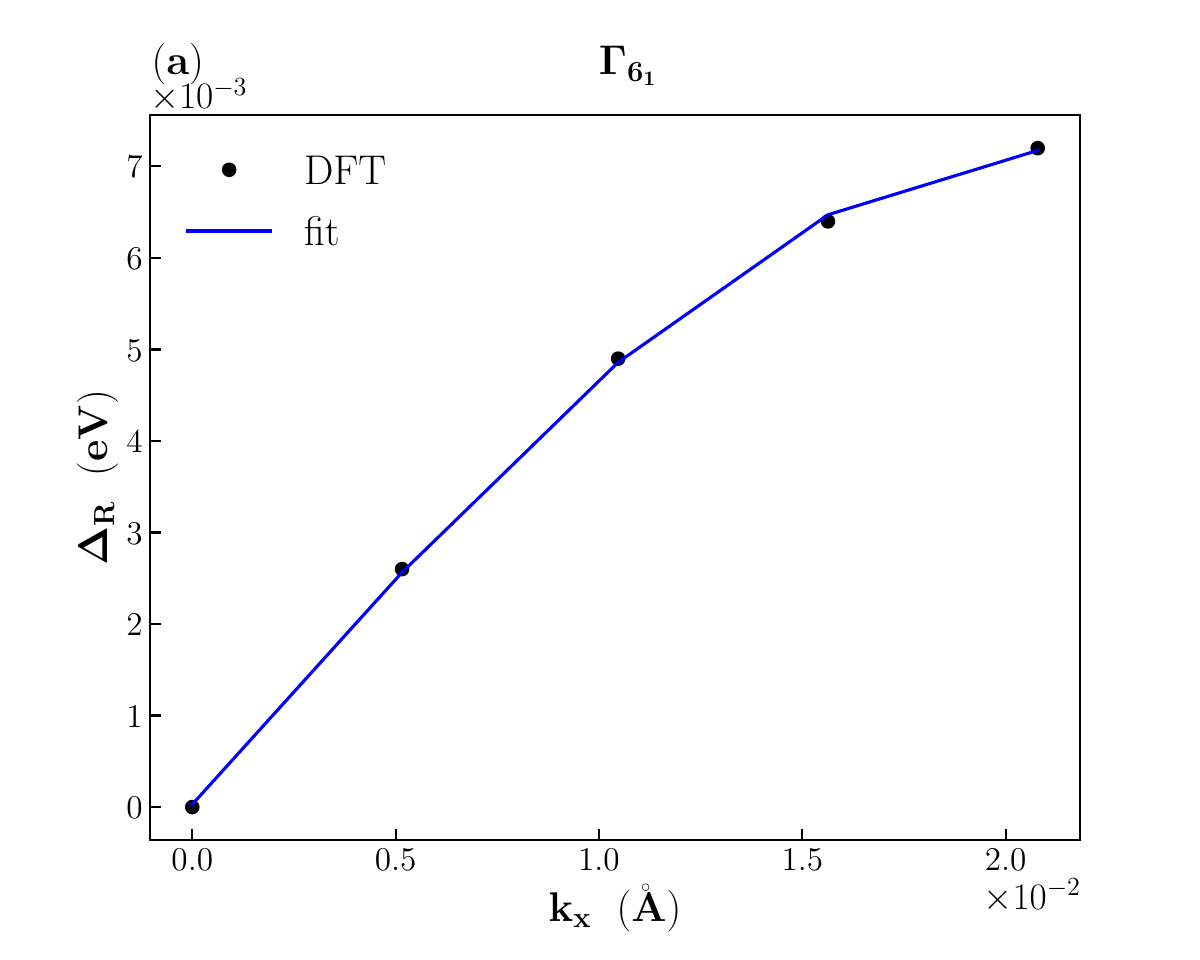}\hspace{-0.5cm}
		\includegraphics[width=0.5\textwidth]{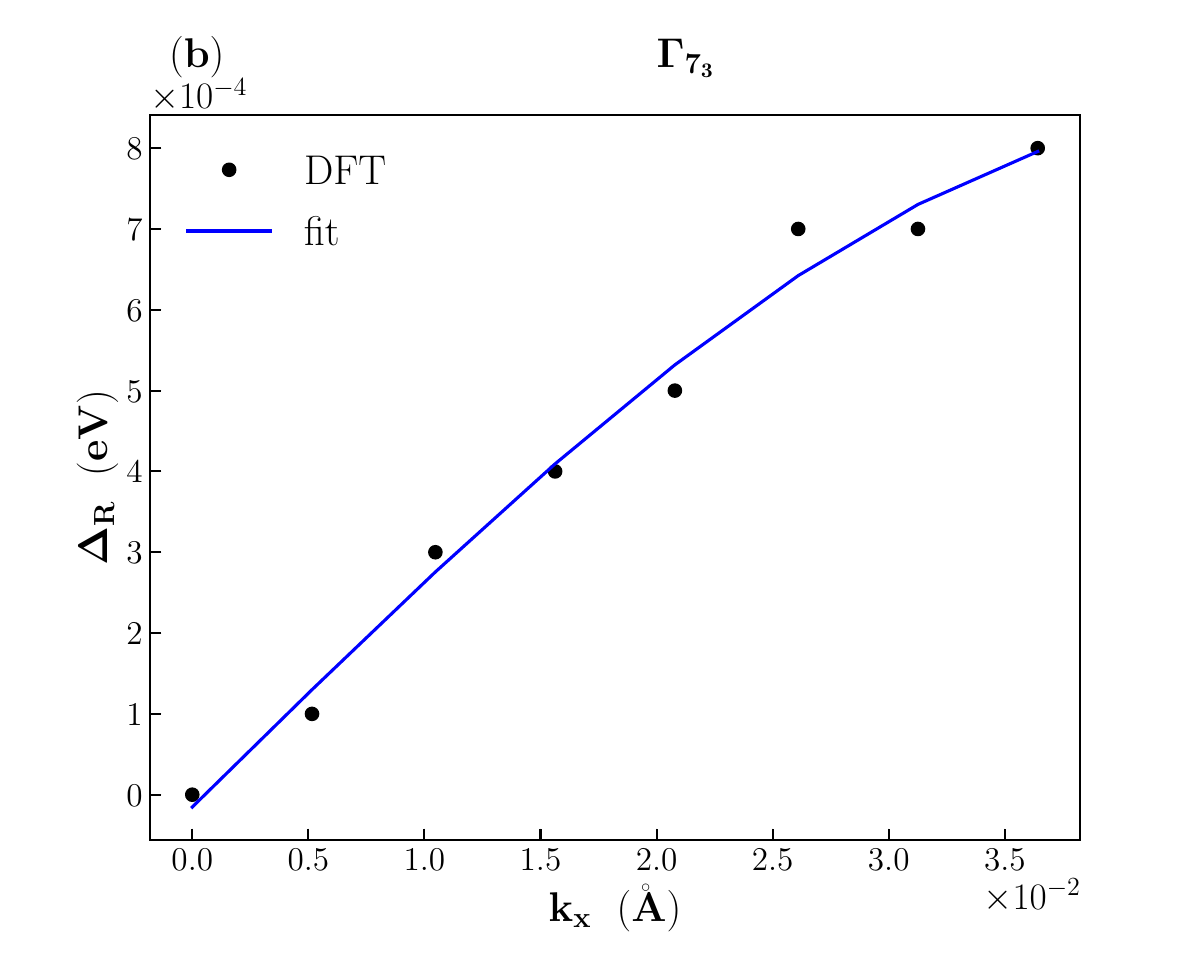}
		\includegraphics[width=0.5\textwidth]{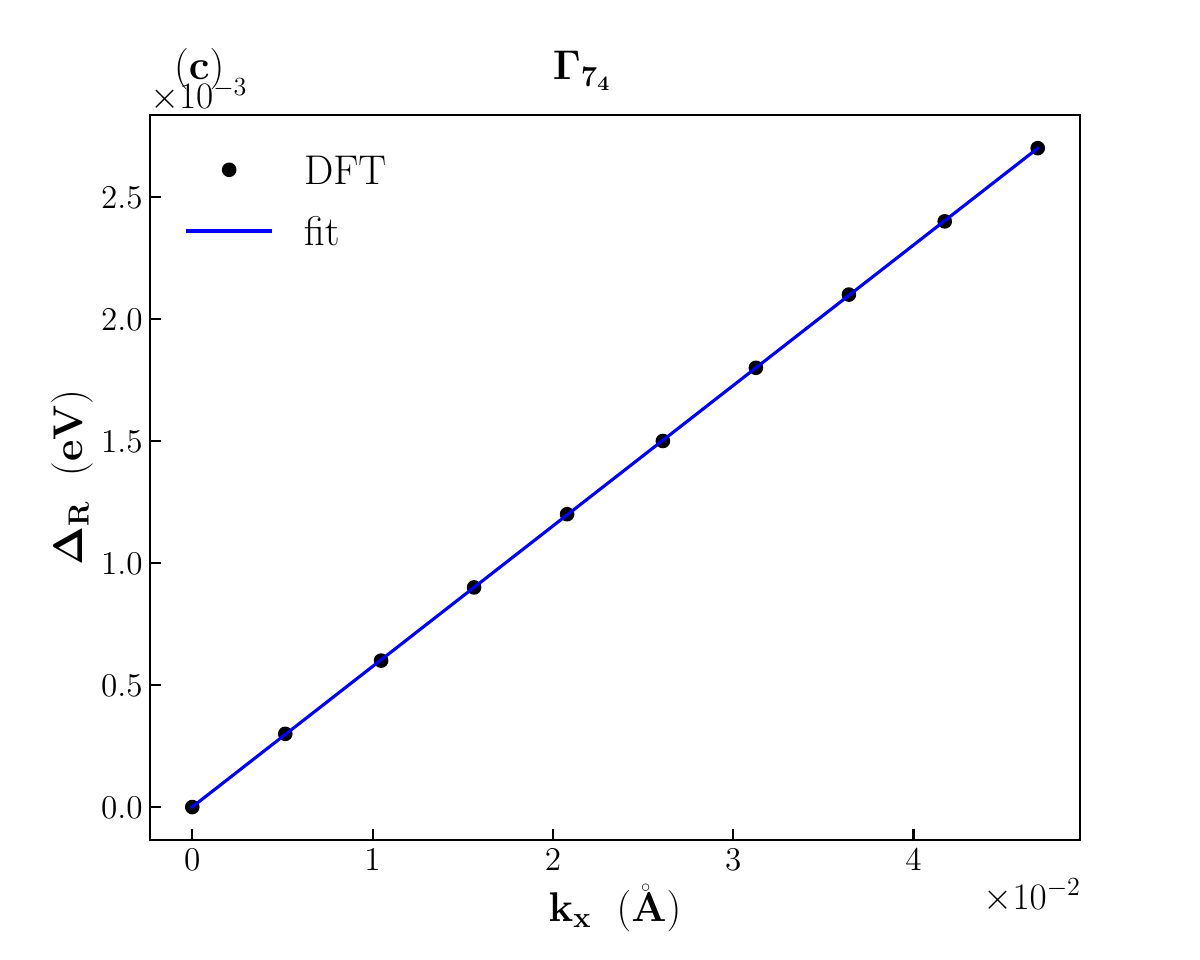}\hspace{-0.5cm}
		\includegraphics[width=0.5\textwidth]{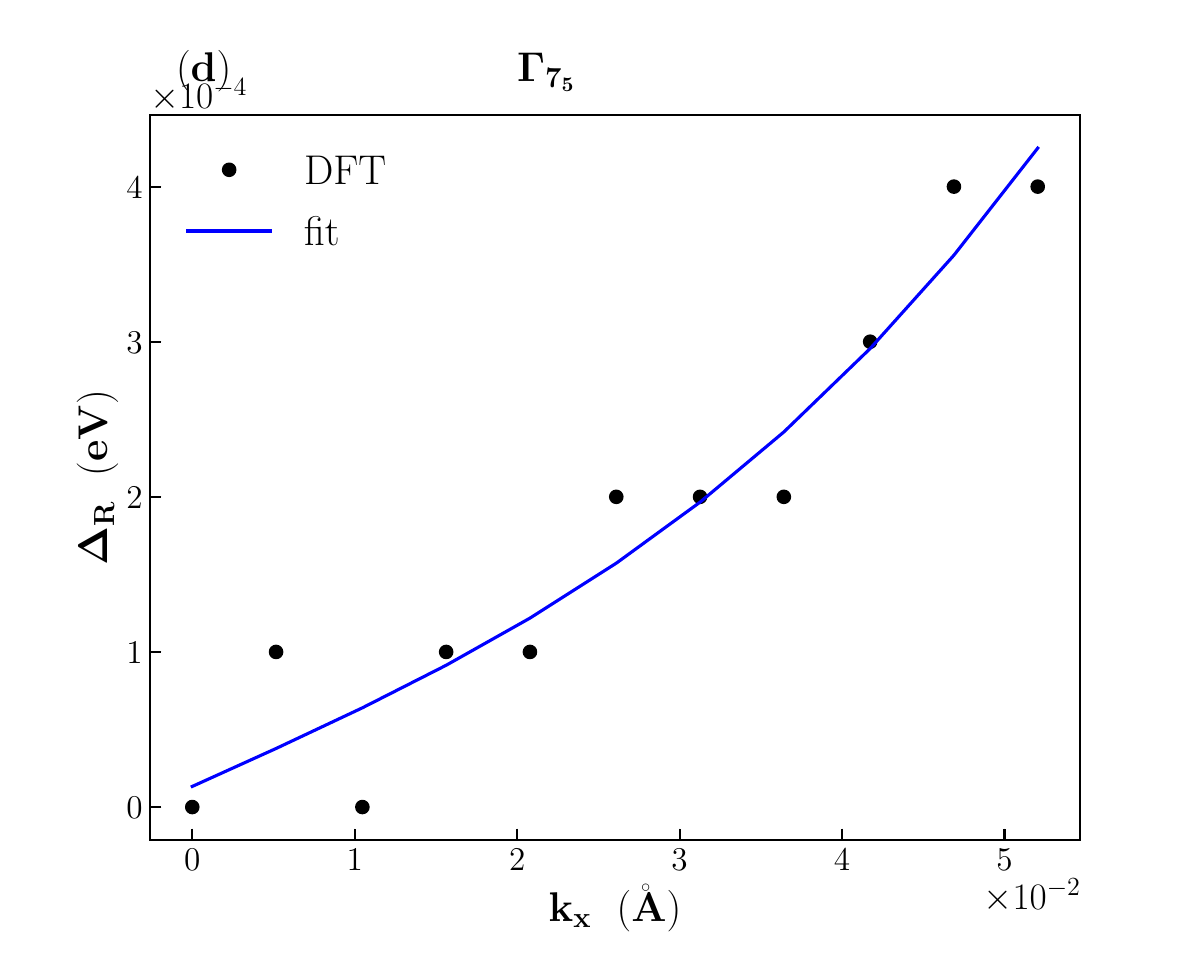}
		\includegraphics[width=0.5\textwidth]{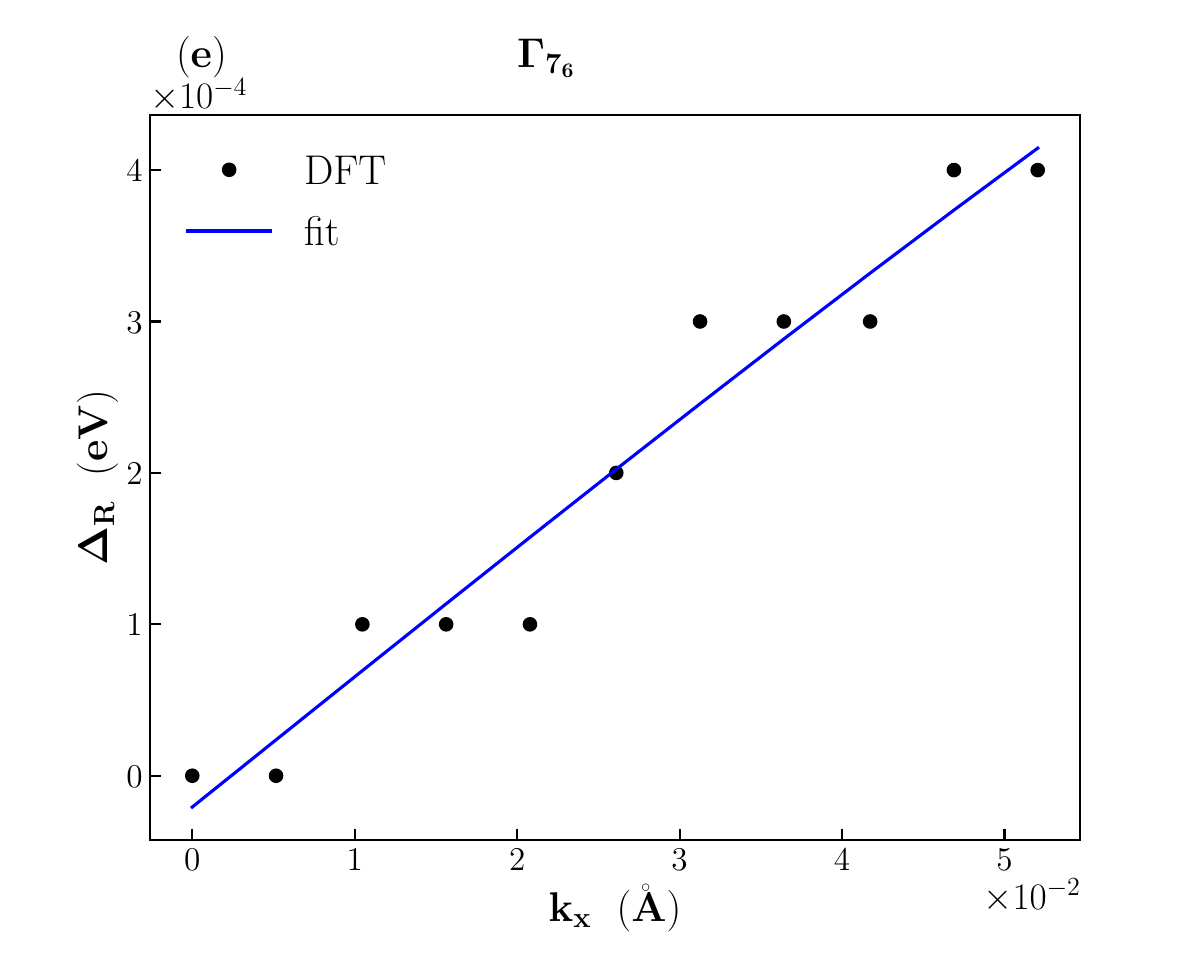}\hspace{-0.5cm}
		\includegraphics[width=0.5\textwidth]{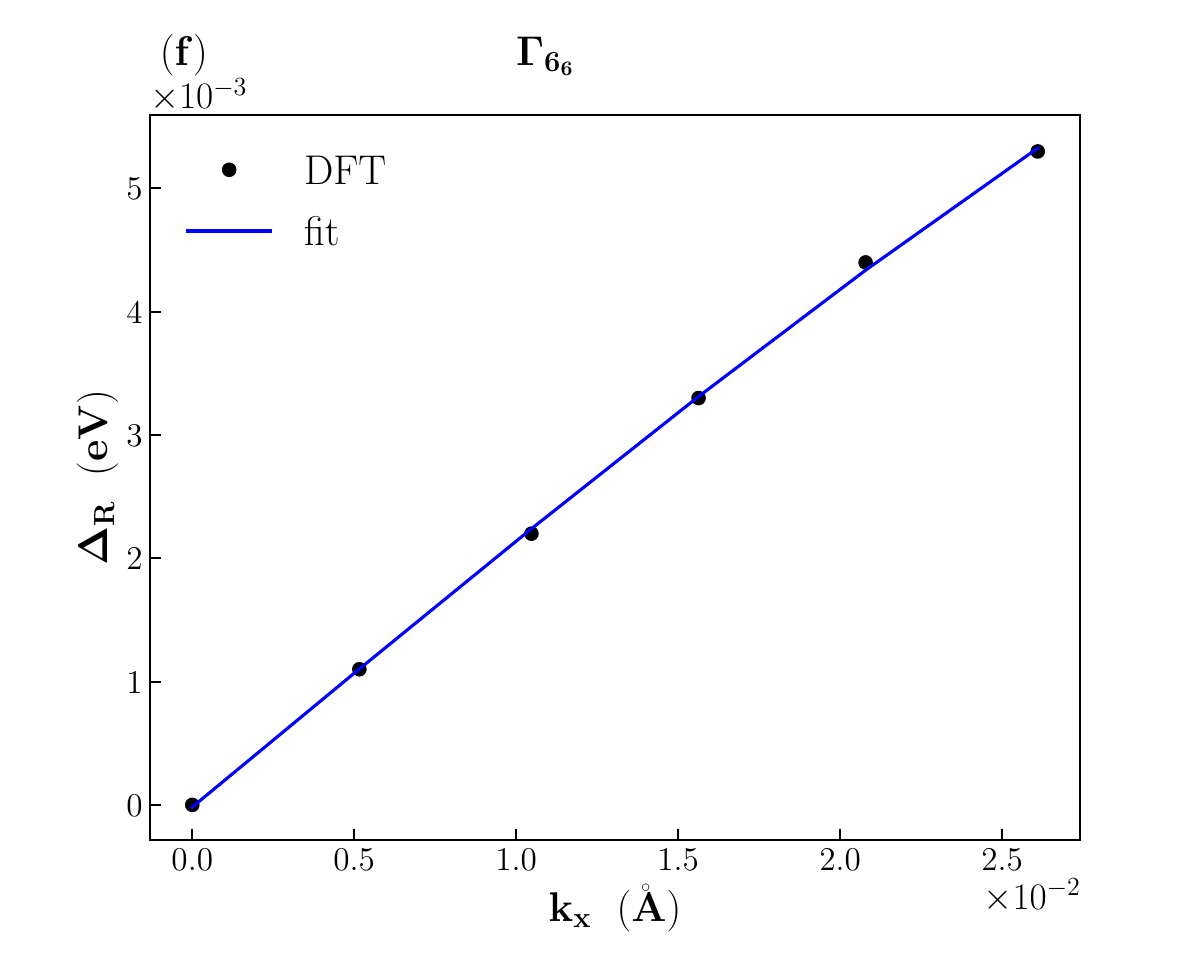}
		\caption{Hamiltonian fitting by using eqn.~\ref{fitting} for
			(a) $\Gamma_{6_1}$
			(b) $\Gamma_{7_3}$ 
			(c) $\Gamma_{7_4}$
			(d) $\Gamma_{7_5}$  
			(e) $\Gamma_{7_6}$ 
			(f) $\Gamma_{6_6}$ bands}
		\label{Fig:pt-fit-sp}
	\end{figure}
	\begin{table}[htbp!]  
		\caption{The table represents the RSO coupling strength for Type-II system obtained by fitting Eqn.~\ref{fitting} with calculated DFT band dispersion data.
			Corresponding plots are shown Fig~\ref{Fig:pt-fit-sp}.} 
		\centering  
		\begin{tabular}{c c c c c c}   
			\hline\hline   
			Orbital  & $\alpha_{R1}$ & $\alpha_{R3}$\\
			(Band Level)&   (eV\AA)    &   (eV\AA$^3$) \\
			
			\hline\hline   
			O-{$\rm{2{p_y}+2{p_z}}$}($ {\Gamma_{6_1}}$) & 0.25 & -182.21 \\  
			Ta-5$\rm{d_{xy}}$ (${\Gamma_{7_3}}$) & 1.41$\times10^{-2}$ & -2.26 \\
			Ta-5$\rm{d_{xy}}$(${\Gamma_{7_4}}$) & 2.88$\times10^{-2}$ & -1.17$\times10^{-2}$ \\
			Ta-5$\rm{d_{xy}}$(${\Gamma_{7_5}}$) & 2.35$\times10^{-3}$ & 0.59 \\
			Ta-5$\rm{d_{xy}}$(${\Gamma_{7_6}}$) & 4.30$\times10^{-3}$ & 4.60$\times10^{-2}$ \\     
			Ta-5$\rm{d_{yz}/d_{zx}}$(${\Gamma_{6_6}}$) & 0.11 & -94.60  \\
			
			\hline
		\end{tabular}  
		\label{tab2r}
	\end{table}
	\section{Type-II heterostructure with fixed layer}\label{t2-wfl}
	\begin{figure}[htbp!]
		\includegraphics[height=0.3\textheight,width=0.5\textwidth,trim={0.5cm 1cm 0.3cm 0.85cm},clip=true]{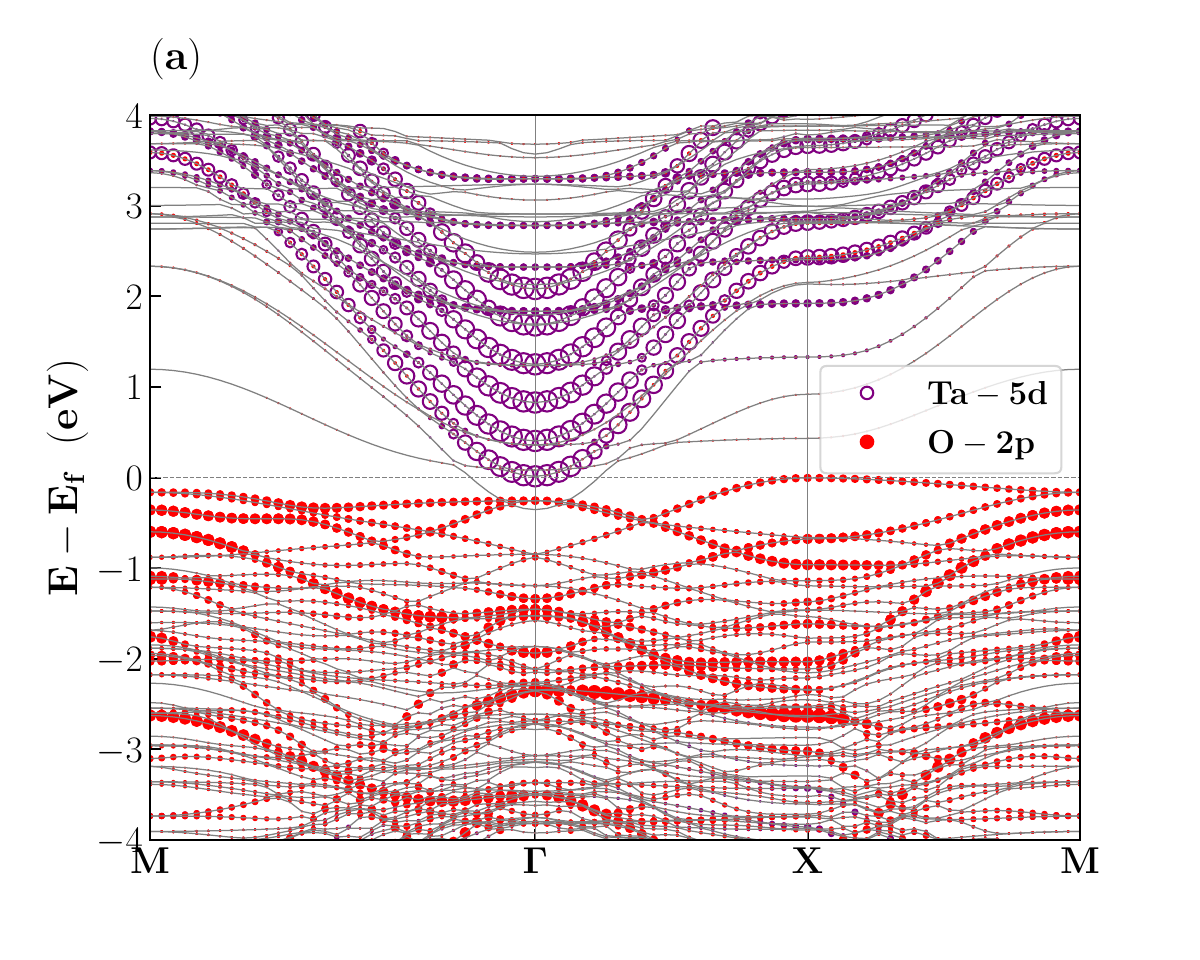}\hspace{-0.6cm}
		\includegraphics[height=0.3\textheight,width=0.5\textwidth,trim={0.5cm 1cm 0.3cm 0.85cm},clip=true]{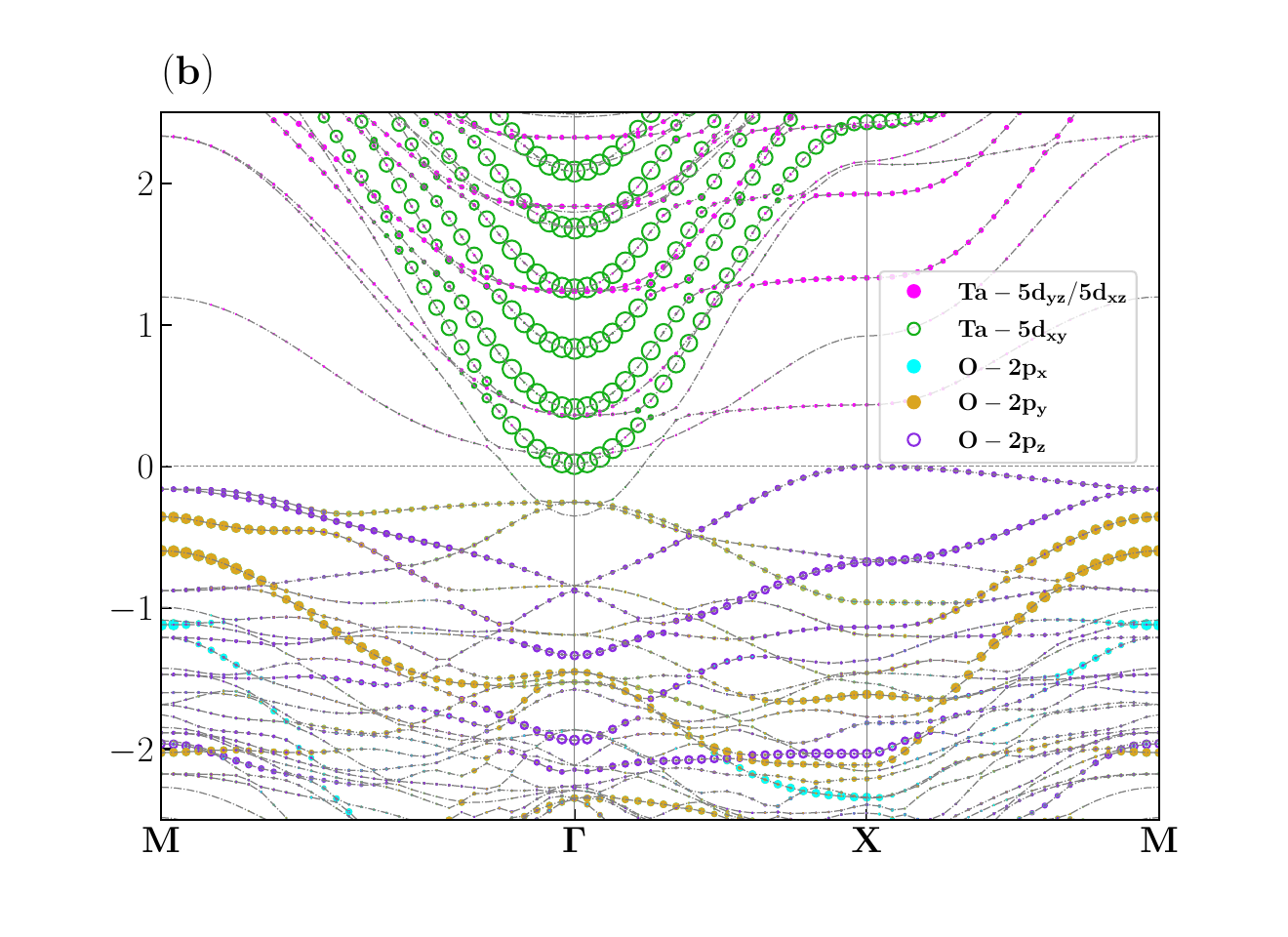}
		\includegraphics[height=0.67\textheight,width=0.55\textwidth,trim={0.3cm 1.4cm 1cm 3cm},clip=true]{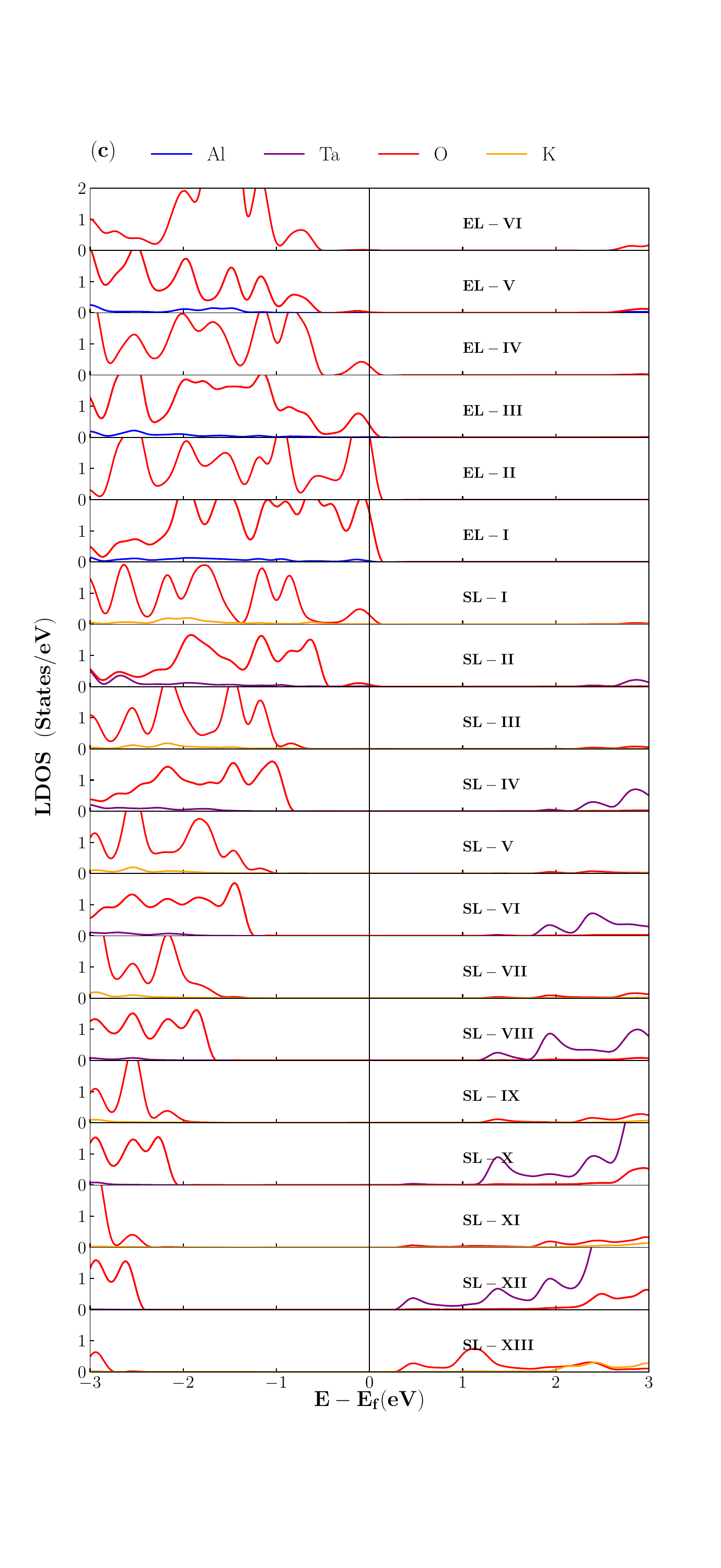}\vspace{-1cm}
		\caption{(a) Band structure of the Type-II heterostructure without SOC. Grey bands represent the states originating from fixed substrate layers, (b) Zoomed-in display of the subbands coming from Ta-5d and O-2p orbitals, and (c) Layer-resolved density of states for the slab system. Fixed substrate layers are not shown in the picture.}
		\label{Fig:ptype-band-rv}
	\end{figure}
	In this section, we have presented the Type-II heterostructure by keeping the last substrate $\rm{TaO_2}$ layer at its fixed position. First we have presented the results without the effect of SOC and it is followed by the results including the effect of SOC.
	Fig.~\ref{Fig:ptype-band-rv}(a) presents the band structure for Type-II system without SOC along $\rm{M}-\Gamma-\rm{X}-\rm{M}$ high symmetry path. 
	O-2p orbitals (marked with solid red circles) are originated from the polar interface created by $\text{KO}^{-}$ and ${\text{AlO}_\text{2}}^{-}$.
	Band structure shows that the conduction bands are consisted of Ta-5d subbands (marked by open purple circle) which arise from $\text{KTaO}_\text{3}$ substrate. 
	The parabolic nature of conduction bands around $\Gamma$ indicates that there is a creation of quantum well, although the 
	conduction band minima does not cross the Fermi level.
	The grey bands without orbital contributions are shown for the fixed layers of the system which do not have any physical significance.
	In Fig.~\ref{Fig:ptype-band-rv}(b) we have presented the individual orbital contributions of Ta-5d and O-2p orbitals in the system.
	It is evident that similar to the Type-I heterostructure, the the degeneracy of the $\rm t_{2g}$ subband of Ta-5d orbital is lifted 
	in an identical manner. Moreover, there is a similar inter-orbital crossing of $\rm d_{xy}$ and $\rm d_{yz}$/$\rm d_{xz}$ bands 
	in this Type-II hetero-interface. In this oxygen-rich interface CFS also plays a crucial role in O-2p orbital splitting.
	After orbital separation, the valence bands become an admixture of $\rm{p_y}+\rm{p_x}$ orbitals and $\rm{p_y}+\rm{p_z}$ orbitals, originated from 
	different epitaxial and substrate sub-layers.
	LRDOS for $\text{KO}^{-}$/$\text{AlO}_\text{2}^{-}$ heterostructure reconfirms the formation of O-rich interface, which presented in 
	Fig.~\ref{Fig:ptype-band-rv}(c). Since both the substrate and the epitaxial layers at the interface are made of negatively charged surfaces, 
	the occurrence of hole gas is expected.  Similar to the type-I system, the last layer of this heterostructure has also 
	been kept fixed, and the density of states (DOS) of the fixed layer has not been shown here. The states of the conduction subbands do not have much contribution at the heterojunction.
	\begin{figure}[htbp!]
		\includegraphics[width=0.5\textwidth,height=0.3\textheight,trim={0.1cm 1cm 1cm 1cm},clip=true]{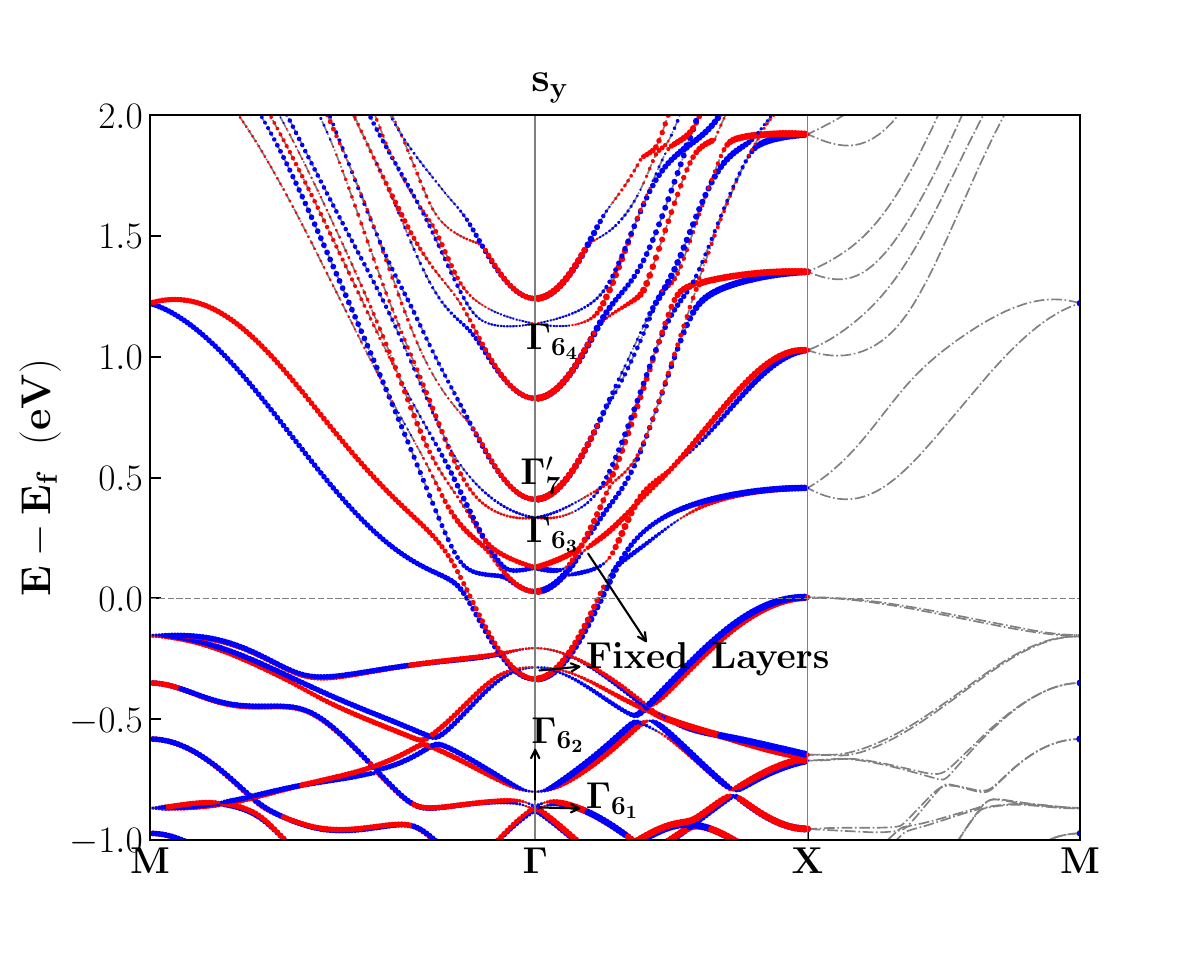}
		\caption{SOC included band structure along $\rm{M}-\Gamma-\rm{X}-\rm{M}$ contributed by Ta-5d and O-2p orbitals. Red and blue colored circles represent up and down spin polarization along $s_y$. The circle size represent the weight of the spin.}
		\label{Fig:pt-soc-band}
	\end{figure}
	\begin{figure}[htbp!]
		\includegraphics[width=0.5\textwidth,height=0.3\textheight,trim={0.05cm 1cm 1cm 0.5cm},clip=true]{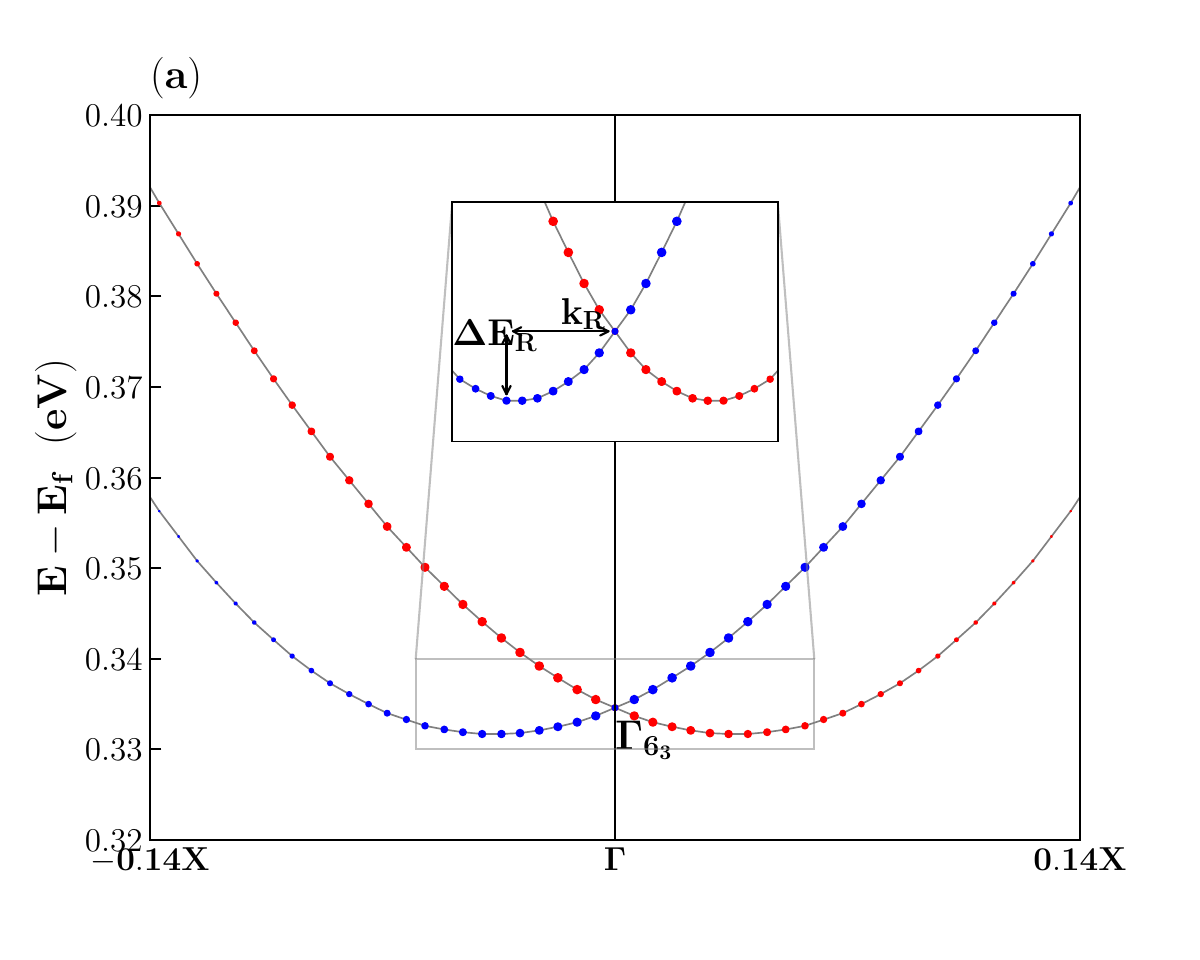}\hspace{-0.7cm}
		\includegraphics[width=0.5\textwidth,height=0.29\textheight,trim={0.05cm 0.65cm 1cm 0.8cm},clip=true]{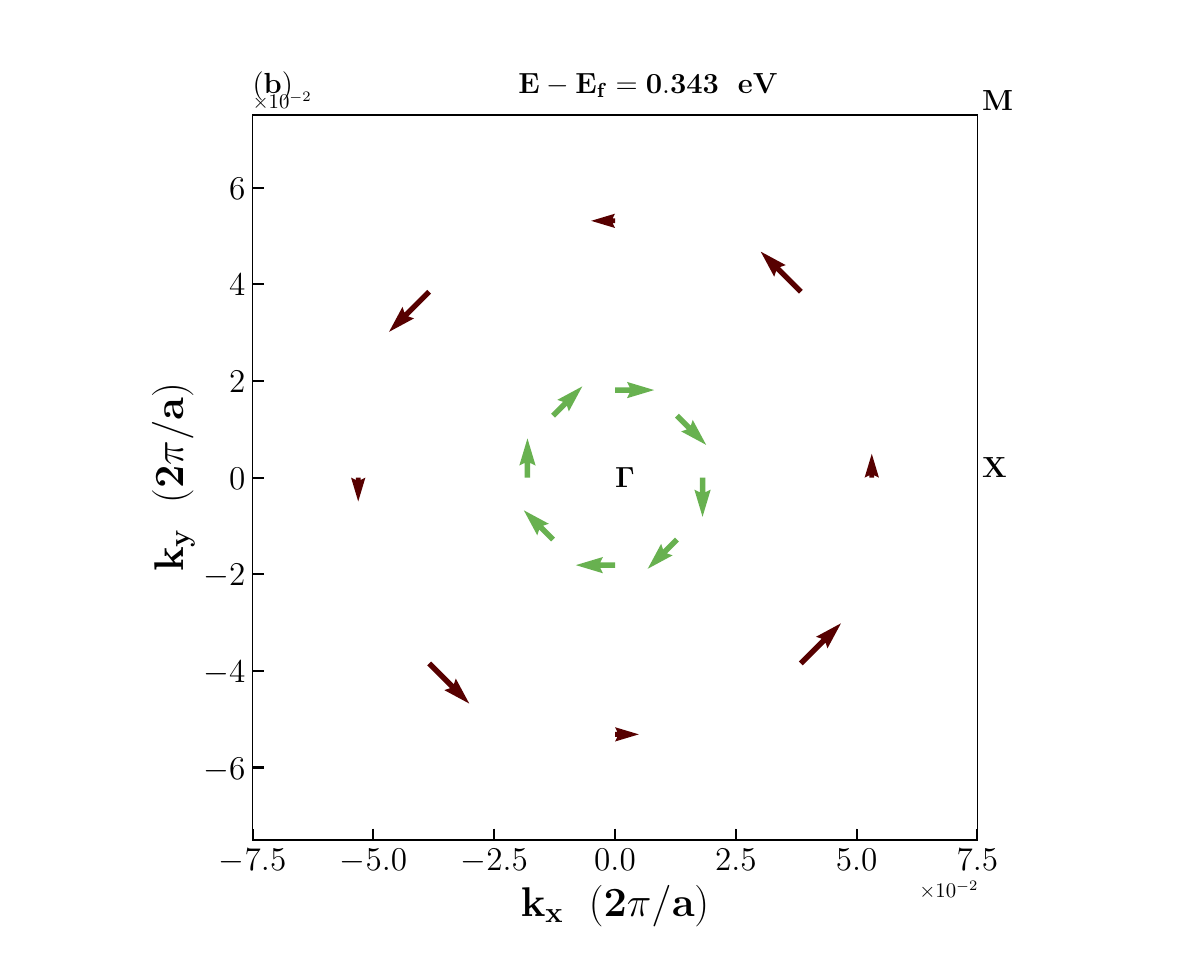}
		\caption{(a) Spin-splitted Ta-$\rm d_{yz}$/$\rm d_{xz}$ band along $\rm{-X}-\Gamma-\rm{X}$ symmetry. $\rm{\Delta{E_{R}}}$ indicates the maximum energy difference between HS and
			the extremum of the parabola at $\rm{k_x}$, where the maximum linear RSO splitting has been found.
			(b) Associated spin-textures of $\Gamma_{6}$ band in the $\rm{k_x-k_y}$ plane at iso-energy $\rm{E-E_{f}}=0.343~eV$.
			Brown and green colors distinguish between the outer and inner bands with opposite spin orientation.}
		\label{Fig:ptype-st}
	\end{figure}
	
	In this section, we have studied Type-II heterostructure in the presence of SOC, and the corresponding band structure is presented in the Fig.~\ref{Fig:pt-soc-band}.
	The presence of SOC in Type-II system leads to the formation of the $\Gamma_{6}$ and $\Gamma_{7}$ levels. 
	The $\Gamma_{6}$ conduction bands consist of the Ta-5d orbitals, while the valence band consist of O-2p orbitals only.  
	Distinct $\Gamma_{6}$ levels are indicated in the band structure by the numerals 1, 2, 3 etc. Unlike Type-I system, here all the 
	$\Gamma_{6}$ levels show RSO spin splitting, whereas $\Gamma_{7}$ levels has no splitting.
	This is an oxygen-rich heterojunction, as was previously mentioned, and O-2p orbitals exhibit RSO-splitting in the valence band region.
	The zoomed in view of the $\Gamma_{6_3}$ band, which is a combination of the $\rm{d_{yz}}$/$\rm{d_{xz}}$ orbitals, is presented in 
	Fig.~\ref{Fig:ptype-st}(a) to demonstrate the k-dependent spin splitting ($\rm{\Delta{E_R}}$) (k path from -0.14X to 0.14X).
	In Fig.~\ref{Fig:ptype-st}, we have presented the spin texture corresponding to the $\Gamma_{6_3}$ band, and the texture 
	clarly shows a modified linear RSO splitting at isoenergy E=$\rm{E_f}+0.343~\textrm{eV}$. The inner band maintains a constant 
	spin projection in the $\rm{k_x-k_y}$ plane, while in the outer band the amplitude of the spin projection periodically changes with momentum. 
	In order to estimate the RSO coupling strength for Type-II system, the DFT band dispersion data is fitted with Eqn.~\ref{fitting}.
	In Table~\ref{tab2}, the results for all the $\Gamma_{6}$ bands are presented.  
	For both the ${\Gamma_{6_1}}$ and $ {\Gamma_{6_2}}$ bands, linear RSO coupling strength is very strong with $\alpha_{R1}\approx780~\textrm{meV\AA}$ and 
	$\alpha_{R1}\approx800~\textrm{meV\AA}$ respectively.
	In the conduction band region, up-to within $\rm{k_x} \lessapprox 0.05~\textrm{\AA}$, ${\Gamma_{6_3}}$ and ${\Gamma_{6_4}}$ level 
	shows linear RSO splitting with coupling strength 200 $\textrm{meV\AA}$ and 260 $\textrm{meV\AA}$ respectively.\\
	\begin{figure}[htbp!]
		\includegraphics[width=0.5\textwidth,height=0.3\textheight,trim={0.05cm 1.2cm 1.2cm 1.2cm},clip=true]{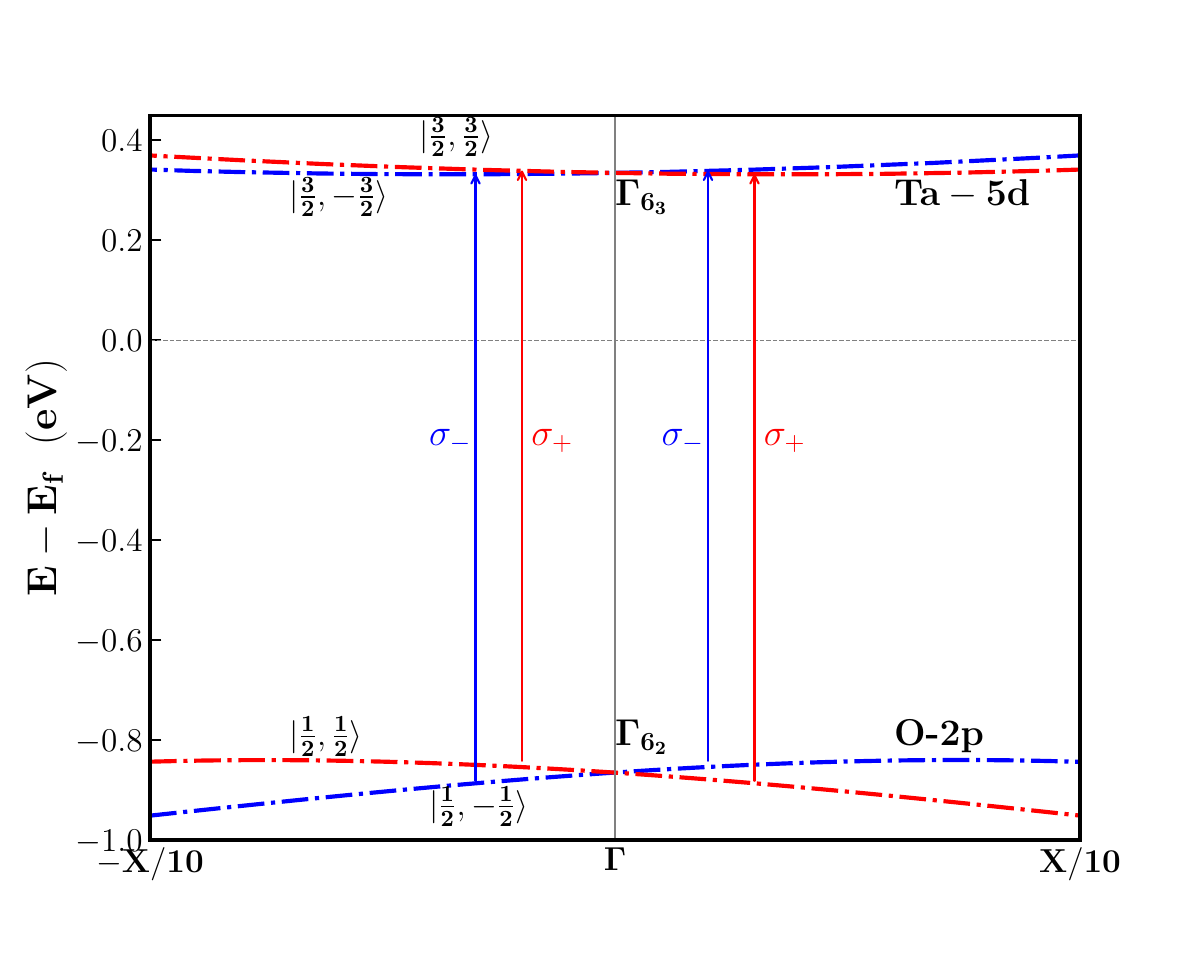}
		\caption{Schematic representation of circularly polarized photogalvanic effect for Type-II system. $\sigma_{+}$ (red) and $\sigma_{-}$ (blue) are the right handed and left handed polarized light. Red and blue arrows show the possible transitions.}
		\label{cpge}
	\end{figure}
	Fig.~\ref{cpge} is a schematic representation of the possible electronic arrangement for producing circularly polarized photocurrent in the LAO/KTO heterostructure. For the right-handed polarized light ($\sigma_{+}$), the interband transition is possible only for $\ket{j,m_j}$=$\ket{{\frac{1}{2}},{\frac{1}{2}}}\rightarrow\ket{{\frac{3}{2}},{\frac{3}{2}}}$ and similarly for left handed polarized light ($\sigma_{-}$) the only possible transition is $\ket{j,m_j}$=$\ket{{\frac{1}{2}},{-\frac{1}{2}}}\rightarrow\ket{{\frac{3}{2}},{-\frac{3}{2}}}$. In the Type-II system, we have shown schematically the transition takes place between ${\Gamma_{6_2}}$ and $ {\Gamma_{6_3}}$ levels which belong to
	EL-I and SL-X.
	
	\subsection{Fitting for RSO parameter in Type-II heterostructure with fixed substrate layer}
	\begin{table}[htbp!]  
		\caption{The table represents the RSO coupling strength for Type-II system obtained by fitting Eqn.~\ref{fitting} with calculated DFT band dispersion data.
			Corresponding plots are shown in supplemental material.} 
		\centering  
		\begin{tabular}{c c c c c c}   
			\hline\hline   
			Band level  & $\alpha_{R1}$ & $\alpha_{R3}$\\
			&   (eV\AA)    &   (eV\AA$^3$) \\
			
			\hline\hline   
			$ {\Gamma_{6_1}}$ & 0.78 & -15.3 \\  
			
			${\Gamma_{6_2}}$ & 0.80 & -19.4  \\
			
			${\Gamma_{6_3}}$ & 0.20 & -3.2 \\
			
			${\Gamma_{6_4}}$ & 0.26 & -85.4 \\
			\hline
		\end{tabular}  
		\label{tab2}
	\end{table} 
	\begin{figure}[t]
		\centering
		\includegraphics[width=0.5\textwidth]{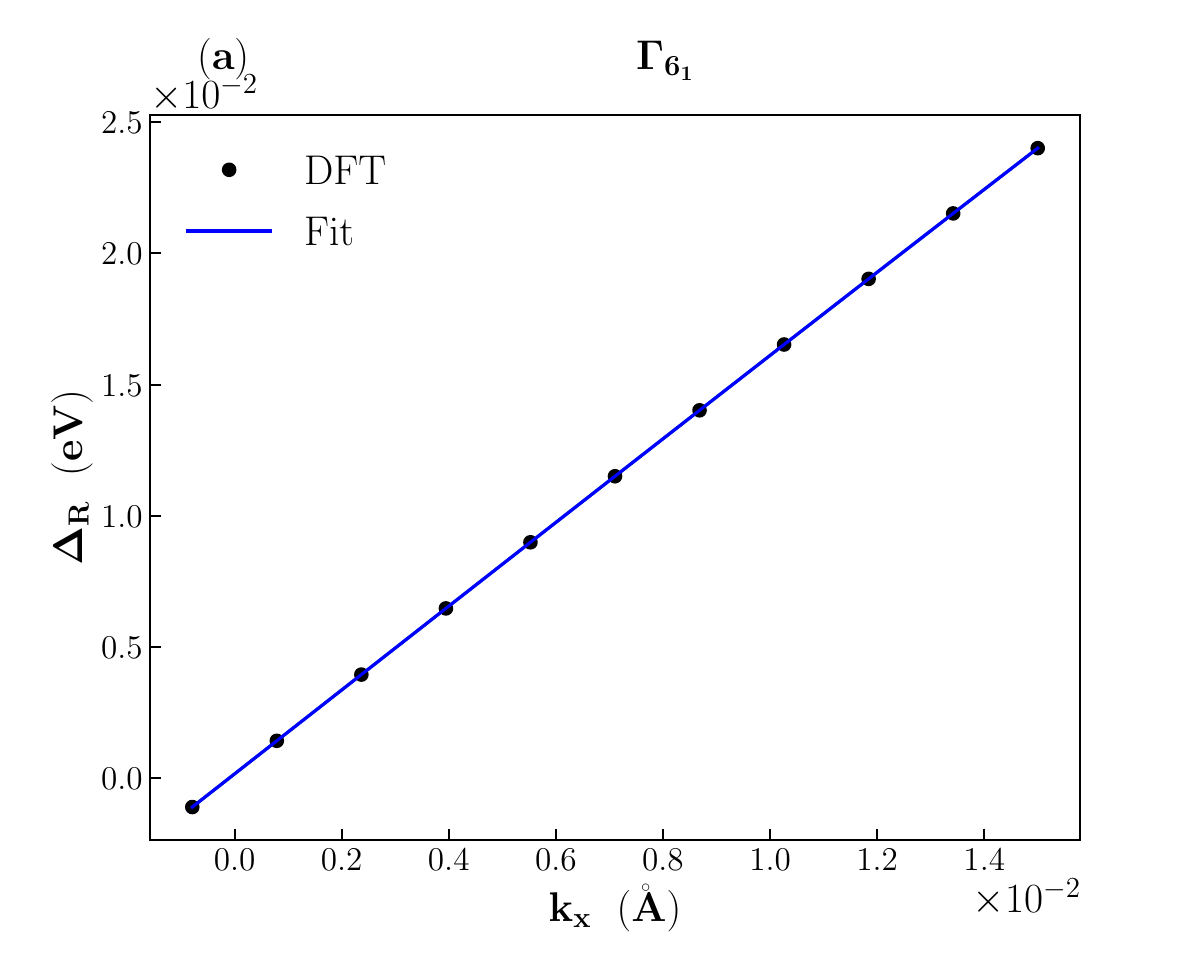}\hspace{-0.5cm}
		\includegraphics[width=0.5\textwidth]{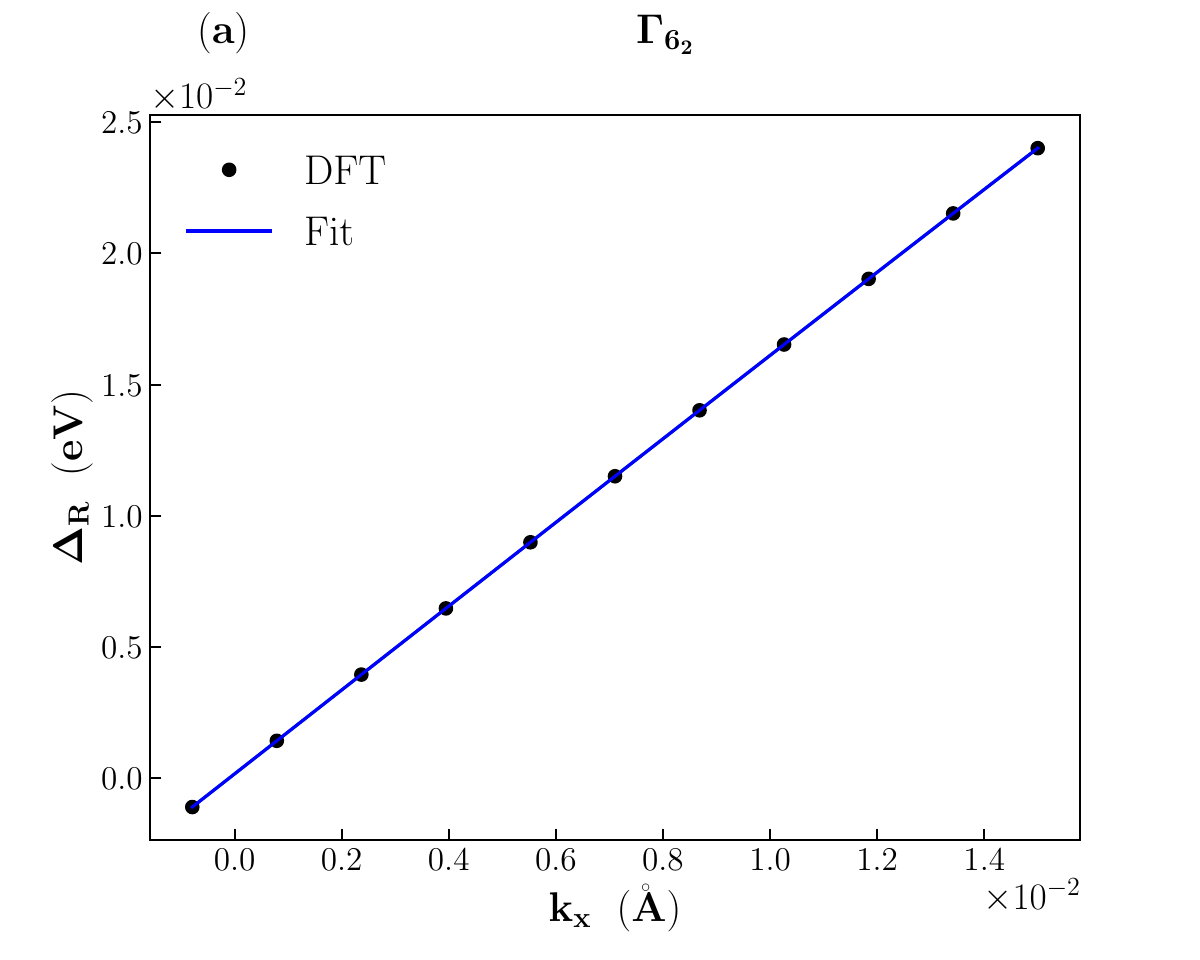}
		\includegraphics[width=0.5\textwidth]{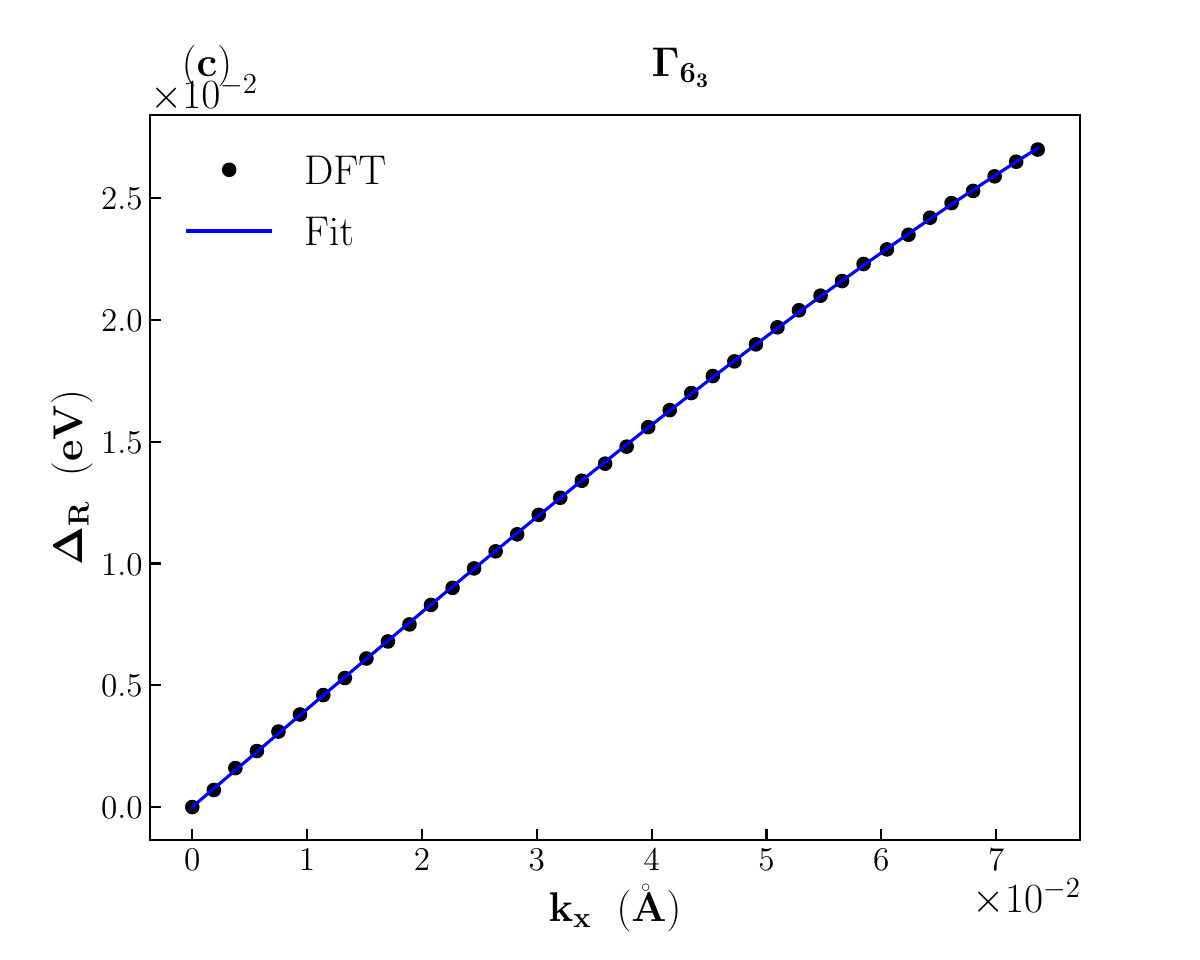}\hspace{-0.5cm}
		\includegraphics[width=0.5\textwidth]{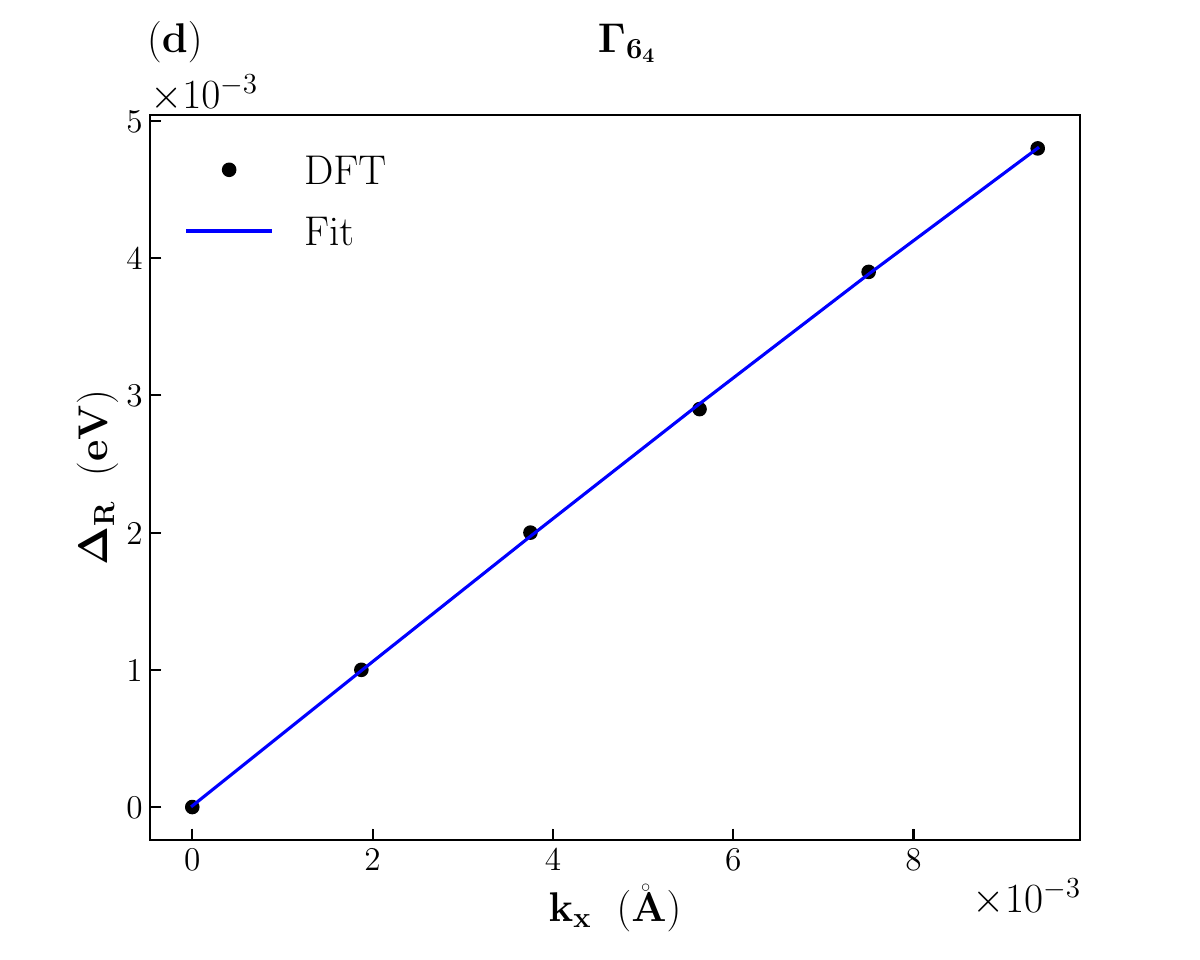}
		\caption{Hamiltonian fitting by using eqn.~\ref{fitting} for
			(a) $\Gamma_{6_1}$
			(b) $\Gamma_{6_2}$ 
			(c) $\Gamma_{6_3}$
			(d) $\Gamma_{6_4}$  bands}
		\label{Fig:ptype-fit-sp}
	\end{figure}
	
\end{document}